\documentclass[numberedappendix]{emulateapj}
\usepackage{graphics}
\usepackage{natbib}
\usepackage{amsmath}
\usepackage{float}
\usepackage{longtable}

\begin{document}
\title{{\bf GARROTXA} cosmological simulations of Milky Way-sized galaxies: General properties, hot gas distribution and missing baryons}
\author{Santi Roca-F\`abrega $^{1}$, Octavio Valenzuela $^{2}$, Pedro Col\'in $^{3}$, Francesca Figueras $^{1}$, Yair Krongold $^{2}$, H\'ector Vel\'azquez $^{4}$, Vladimir Avila-Reese $^{2}$ and Hector Ibarra-Medel $^{2}$}
\affil{$^{1}$ Departament d'Astronomia i Meteorologia and IEEC-UB, Institut de Ci\`encies del Cosmos de la Universitat de Barcelona,\\
     Mart\'i i Franqu\`es, 1, E-08028 Barcelona.\\
$^{2}$ Instituto de Astronom\'ia, Universidad Nacional Aut\'onoma de M\'exico, A.P. 70-264, 04510, M\'exico, D.F.; Ciudad Universitaria, D.F., M\'exico.\\
$^{3}$ Instituto de Radioastronom\'ia y Astrof\'isica, Universidad Nacional Aut\'onoma de M\'exico, A.P. 72-3 (Xangari), Morelia, Michoac\'an 58089, M\'exico.\\
$^{4}$ Instituto de Astronom\'ia, Universidad Nacional Aut\'onoma de M\'exico, A.P. 877, 22800, Ensenada, M\'exico.\\
}
\begin{abstract}
We introduce a new set of  simulations  of Milky Way-sized galaxies  using the AMR code ART + hydrodynamics  in a $\Lambda$CDM  cosmogony.
The simulation series is  named GARROTXA and follow the formation of a halo/galaxy  from z~$=$~60 to z~$=$~0.
The final virial mass of the system is $\sim$7.4$\times$10$^{11}$M$_{\odot}$. 
Our results are as follows: (a) contrary to many
previous studies, the circular velocity curve shows no central peak and overall agrees
with recent MW observations. (b) Other quantities, such as M$\_{*}$(6$\times$10$^{10}$M$_\odot$) and R$_d$
(2.56 kpc), fall well inside the observational MW range. (c) We measure the disk-to-total ratio kinematically and find 
that D/T=0.42. (d) The cold gas fraction and star formation rate 
(SFR) at z=0, on the other hand, fall short from the values estimated for the Milky Way.\\
As a first scientific exploitation of the simulation series, we study the spatial distribution of 
the hot X-ray luminous gas. We have found that most of this X-ray emitting gas is in a halo-like distribution accounting for an important fraction but not all of the missing baryons. An important amount of hot gas is also present in filaments. In all our models there is not a massive disk-like hot gas distribution 
dominating the column density. Our analysis of hot gas mock observations reveals that 
the  homogeneity assumption leads to an overestimation of the total mass by factors  3 to 5 or to an underestimation by factors $0.7-0.1$, depending on the used observational method. Finally, we confirm a clear correlation between the total hot gas mass and the dark matter halo mass of galactic systems.
\end{abstract}
\section{Introduction}\label{sec:intro}
One of the most challenging problems 
that hydrodynamical simulations have
faced since the pioneering works of \citet{Evrard1988,Hernquist1989,Cen1990,Navarro&Benz1991,Navarro1993} has been to produce,
within the standard $\Lambda$CDM hierarchical structure formation scenario, systems that 
look like real disk galaxies;
that is, galaxies with extended disks and present-day star formation rates, disk-to-bulge ratios, and 
gas and baryonic fractions that agree with observations.

It has been long known that disks should form when gas cools and condenses within dark matter halos 
\citep{White&Rees1978,Fall&Efstathiou1980,MoMao&White1998}
conserving angular momentum obtained through external torques 
from neighbouring structures \citep{Hoyle1953,Peebles1969}.
Yet, the first halo/galaxy hydrodynamical simulations \citep{Navarro&Benz1991, Navarro&Steinmetz2000}
inevitably ended up with small and compact disks and massive spheroids that dominated the 
mass. Two effects were found to be the responsible of this result. First, artificial losses of angular momentum caused by 
insufficient resolution and other numerical effects \citep{Abadi2003,Okamoto2003,Governato2004,Kaufmann2007}. 
Second, the dynamical friction
that transfers the orbital angular momentum of merging substructures to the outer halo and causes cold baryons 
(stars plus gas) to sink to the center
of the proto-galaxy \citep[e.g. ][]{Hernquist1995}. Once in the center, baryons that are still in the form of gas
are quickly converted to stars. These newborn stars will join the stellar component and both will form 
an old spheroid, leaving no gas for the formation of new disk stars at latter times \citep[e.g.][]{Maller&Dekel2002}.
Later, the increase in resolution in the numerical
simulations led to several authors \citep[e.g.][]{Robertson2004,Okamoto2005,Governato2004} to succeed 
in getting more realistic disk galaxies. This better resolution was achieved using the so called zoom-in technique. 
This technique allows simulators to obtain high resolution by choosing a small region, almost always
a sphere centered on a specific halo, from a relatively big box, and rerun it with higher
resolution \citep[e.g. ][]{Klypin2001}. To obtain 
a ``MW-like'' system the authors selected halos with masses similar to the one estimated for
the MW ($\sim$10$^{12}$~M~$_{\odot}$) and halos that have not suffered a major merger since 
$z$~$\sim$~1.5 \citep[e.g.][]{Robertson2004,Governato2004,Okamoto2005,Scannapieco2009,Guedes2011,Moster2014,Vogelsberger2014}.

The recent relative success on forming more realistic disk galaxies has not been limited to resolution.
The implementation of new subgrid physics such as an efficient supernovae (SNe) feedback 
\citep{Stinson2006} and new star formation recipes have shown to be important
\citep{Gnedin2009}.
The more effective stellar feedback acts by removing low-angular-momentum material 
from the central part of (proto)galaxies \citep[e.g.][]{Brook2012} while a SF based
on the molecular hydrogen abundance is more realistic.
These improvements make simulated galaxies to follow observed scaling relations, 
like the Tully-Fisher, for the first time
 \citep{Governato2007} and thus relax the tension that existed between observations and $\Lambda$CDM predictions 
\citep{Governato2010}. In galaxy formation simulations, gas is accreted in filamentary cold flows that never 
is shock-heated to the 
halo virial temperature \citep{Keres2005,Dekel2009}. This cold gas is the fuel for the late star formation in disks 
\citep{Keres2009,Ceverino2010}. For instance, authors like 
\citet{Scannapieco2009,Stinson2010,Piontek&Steinmetz2011,Agertz2011,Brooks2011,Feldmann2011,Guedes2011}
obtained realistic rotationally supported disks with some properties closely resembling the MW ones.

Until recently, simulated disk galaxies with rotation curves similar to that of the Milky Way were
scarce. Most works used to obtain systems with too peaked and declining rotation curves
\citep[e.g. ][]{Scannapieco2012,Hummels2012}. \citet{Agertz2011} showed that by fine tuning
the values of the SF efficiency and the gas density SF threshold parameters 
it was possible to avoid the formation of galaxies with such peaked rotation
curves. 
In more recent high resolution simulations, including ours, this is not
longer a problem \citep{Aumer2013,Mollitor2014,Marinacci2014,Murante2015,Agertz2015,Keller2015,SantosSantos2016},
in part due to an added ``early-feedback''  \citep{Stinson2013}
to the thermal SNe one\footnote{These authors need to delay or switch off cooling for
some time once the ``regular'' SN feedback is switch on to avoid overcooling.}, the implementation of an efficient kinetic feedback, and/or the inclusion of the AGN feedback. The early feedback is attributed to a feedback that is present from the very beginning, before the first supernova explodes, and it may include radiation pressure \citep{Krumholz2015}, photoheating of the gas due to the ionizing radiation of massive stars \citep{TrujilloGomez2015}, stellar winds, etc. Currently the need of such effects are widely acknowledged but their correct implementations are still a challenge \citep{Oman2015,GonzalezSamaniego2014}.  

Obtaining systems that resemble the MW open several possibilities on the study of galaxy formation and evolution. For instance,
it is interesting to study the origin of the galactic halo coronal gas, their properties and how they can account for a fraction of the missing baryons.
The lack of baryons in the universe was reported by \citet{Fukugita1998} when they found that cosmological baryon fraction inferred 
from Big Bang nucleosynthesis is much higher than the one obtained by counting baryons at redshift z~=~0. This problem was confirmed
 when a more accurate value for the cosmological baryon fraction was obtained from the high precision data of WMAP 
\citep{Dunkley2009} and Planck \citep{Planck2013}. In fact, the cosmic ratio $\Omega_b$/$\Omega_M$ is 3$-$10 
times larger than the one observed for galaxies. To solve this missing baryons problem, following the natural prediction from structure formation models \citep{White&Rees1978,White&Frenk1991}, authors like \citet{Cen&Ostriker1999,Dave2001,Bregman2007} proposed that galactic winds, SNe feedback or strong AGN winds ejected part of the galactic baryons to the dark matter halo and circumgalactic medium (CGM) as hot gas.  Others authors like \citet{Mo&Mao2004} proposed that also part of the gas never collapsed into the dark matter halos as it was previously heated by SNe of Population III.
From the point of view of large volume hydrodynamical galaxy formation simulations, some attempts 
have been made in order to investigate the presence and detectability
of hot gas halo corona. \citet{Toft2002} presented one of the first studies towards this direction, however due to the inefficient feedback, the amount of hot gas in their simulated halo was too small. Global properties of hot gas halo corona in MW-sized simulations have also been studied \citep[e.g.][]{Guedes2011, Mollitor2014}. More recently, models that account for the SNe feedback and implement new star formation processes, succeeded in obtaining more realistic
results when  compared with observations \citep{Crain2010,Crain2013}.

From an observational point of view the hot gas halo corona was proposed to be detected by its X-ray emission. 
First detections where done by \citet{Forman1985} in external galaxies and later
 on several authors get better detections in other systems and studied them deeper \citep[e.g.][]{Mathews&Brighenti2003,Li2008,Anderson&Bregman2010,
Bogdan&Gilfanov2011}. More recently, using Chandra, XMM-Newton, FUSE and other instruments \citep[e.g. ][]{Hagihara2010,Hagihara2011,Miller2013} it was detected a large reservoir of this hot gas surrounding our galaxy \citep[e.g. $\sim$6$\times$10$^{10}$~M~$_{\odot}$ in][]{Gupta2012} and it was proposed that it could account for a fraction of the so called missing baryons mass. These detections were done by the analysis of X-ray absorption lines from OVII and OVIII which only exist in environments with temperatures between 10$^6$ and 10$^7$~K, the MW halo has a temperature of about $log$~T~=~6.1$-$6.4 \citep{Yao&Wang2007,Hagihara2010}. The problem of using this technique is that these X-ray absorption lines only can be observed in the directions of extragalactic luminous sources (QSO, AGN, ...)
or galactic X-ray emitters as X-ray binaries. This observational strategy gives a limited information about hot gas distribution and position,
inside the galactic halo or in the intergalactic medium. Using this limited information, several recent works \citep{Gupta2012,Miller&Bregman2013,Gupta2014} obtained a value for the total hot gas mass embedded in the galactic halo that accounts for about 10$-$50$\%$  of missing baryons. To obtain these total hot gas mass from the low number of available observations the authors needed to assume a simple hot gas density profile. It is important to be aware that this assumption can lead to biased total hot gas mass values. Recently \citet{Faerman2016} used an analytic phenomenological model to study the warm/hot gaseous coronae distribution in galaxies. They 
have used the MW hot gas X-ray and UV observations as input for their model and found that the warm/hot halo coronae may contain a large reservoir of gas. They conclude that if metallicity is of about 0.5 solar it can account for an important fraction of the missing baryons.\\

Here, we introduce a new set of MW-sized simulations with a high
number of DM particles and cells ($\sim$7$\times$10$^6$). On the
other hand, the spatial resolution, the side of the cell in the
maximum level of refinement, is 109~pc. {\it These are highly resolved
simulations even for today standards}. The
simulated galaxy ends up with an extended disk and an unpeaked slowly 
decreasing rotation curve. It is important to mention that the subgrid physics used in the
present work was shown to produce realistic low-mass galaxies, at
least as far as the rotation curves was concerned \citep[e.g. ][]{Colin2013}. 
In the GARROTXA series of simulations the temperature reached by the gas in the cells, where
stellar particles are born, is actually higher than the one obtained in, 
for example, the simulations by \citet{GonzalezSamaniego2014} because the SF efficiency
factor is higher (0.65 versus 0.5). This along with the fact that here
the gas density threshold is lower, 1~cm~$^{-3}$ instead of 6~cm~$^{-3}$, makes
the stellar feedback in GARROTXA simulations stronger.\\
Using our simulations we have also studied how the hot gas component embedded in the DM halo can account for part of the missing 
baryons in galaxies. 
 A novelty of the present paper is the determination, in our GARROTXA series
of simulations, of the hot halo gas distribution in a full sky view which
offers us an opportunity to detect {\it observational biases} in the
determination of its mass, when only a small number of line of sight
observations are used.\\

This paper is laid out as follows. In Sec.~\ref{sec:sim} we present the code and initial conditions we have used. 
In Sec.~\ref{sec:results} we introduce the general properties of our MW-sized simulations, 
whilst in Sec.~\ref{sec:results1} we present the study of their hot gas component. We summarize our conclusions in Sec.~\ref{sec:conclusions}.

\section{GARROTXA simulation}\label{sec:sim}

\subsection{The code}

The numerical simulations we introduce in this work (GARROTXA: GAlaxy high Resolution Runs in a cOsmological ConteXt using ART) have been computed 
using the Eulerian hydrodynamics + N-body ART code \citep{Kravtsov1997,Kravtsov2003}.
This is the Fortran version of the code introduced in \citet{Colin2010}
in the context of the formation of low-mass galaxies. This version differs from
the one used by \citet{Ceverino2009}, among other things, in the
feedback recipe; unlike them, we deposit {\it all} the energy coming from stellar
winds and supernovae in the gas {\it immediately after} the birth of the stellar 
particle. This sudden injection of energy is able to raise the temperature of the
gas in the cell to several 10$^7$~K, high enough to make the cooling time
larger than the crossing time and thus avoiding most of the overcooling\footnote{
This temperature is the one that the cell acquires if one assumes that all
the energy injected by the SNe (and stellar winds) is in the form of thermal
energy. Its exact value will depend on the assumed IMF and, $\ filon_{SF}$ and SNe
energy values.}. 
Here we have used it to obtain high resolution Milky Way sized galaxies.\\

The code is based on the adaptive mesh refinement technique which
allows to increase the resolution selectively in a specified region of interest around
a selected dark matter halo.
The physical processes included in this code are the cooling of the gas and its subsequent conversion into stars,
the thermal stellar feedback, the self-consistent advection of metals, and a UV heating background source.
 The used total cooling and heating  rates incorporate Compton heating/cooling,
 atomic, and molecular hydrogen and metal-line cooling, UV heating from a cosmological background radiation \citep{Haardt&Madau1996}, and are
 tabulated for a temperature range of 10$^2$ K $<$ T $<$ 10$^9$ K and a grid of densities, metallicities (from log(Z) = -3.0 to log(Z)=1.0, where Z is in solar units)
, and redshifts, using the CLOUDY code \citep[][version 96b4]{Ferland1998}.

The star formation (SF) is modeled as taking place in the coldest and densest collapsed regions, defined by T~$<$~T$_{SF}$ and n$_g$~$>$~n$_{SF}$,
 where T and n$_g$ are the temperature
 and number density of gas, respectively, and T$_{SF}$ and n$_{SF}$ are the temperature and density SF threshold, respectively. A stellar particle
 of mass m$_∗$ is placed in all grid cells where these conditions are simultaneously satisfied, and this mass is removed from the gas mass in the cell.
The particle subsequently follows N-body dynamics. No other criteria are imposed. In most of simulations presented here, the stellar particle
 mass, m$_∗$, is calculated by assuming that a given fraction (SF local efficiency factor $\epsilon_{SF}$) of the cell gas mass, m$_g$, is converted
 into stars; that is, m$_∗$~=~$\epsilon_{SF}$~m$_g$, where $\epsilon_{SF}$ is treated as a free parameter. In MW-sized models presented here
we have used $\epsilon_{SF}$~=~0.65, T$_{SF}$~=~9000~K and n$_{SF}$~=~1~cm$^{-3}$, where this last is the density threshold in hydrogen atoms per cubic centimeter. As shown in \citet{Colin2010}, these values successfully reproduce realistic low-mass galaxies at least as the circular velocity was concerned. In \citet{Colin2010} the 
authors also found that the reduction of $\epsilon_{SF}$ makes the temperature reached by the gas in the star forming cell to be too low , as a consequence the cooling became too efficient. They also show that the strength of the stellar feedback recipe used here depends also on the value of n$_{SF}$: the lower its value, the stronger the effect of the feedback. We tested that most of the overcooling is avoided when n$_{SF}$ is around 1~cm$^{-3}$ and $\epsilon_{SF}$~=~0.65. Finally, the value T$_{SF}$ is almost irrelevant as long it is set below or close to 10$^4$ K because this is always achieved when the density threshold is reached. Incidentally, the Jeans length of an isothermal gas, with a density of n$_H$=3 cm$^{-3}$ and a temperature of 3000 K  is 638 pc, which is more than four times the length of the cell at the highest refinement level at present-day \footnote{As n$_{SF}$ is 1 cm$^{-3}$, it is expected that
at the moment of the stellar particle formation the gas density to be greater than 1 but not much greater. Moreover, as these simulations do not have self-shielding the average gas temperature in the disk lies around the few thousands degrees.}  (109 pc). This value has been obtained using the standard perturbative derivation of Jeans criteria \citep{Kolb1990,Shu1991,Binney2008} that delivers the expression shown in Eq.\ref{eq:1} for the Jeans length, where c$_s$ is the sound speed and the $\rho_0$ the mass density of the star forming gas (cold and dense gas). In Appendix B we show a histogram of the Jeans length, in units of the corresponding cell, for the cold gas.\\

\begin{eqnarray}
\label{eq:1}
\lambda_J=\sqrt{\frac{\pi c_s^2}{G\rho_0}}
\end{eqnarray}

In this work we have found that these values also successfully  reproduce a galaxy rotation curve similar to the one of the Milky Way at z=0 for galaxies with mass similar to the one of the MW. 
Since stellar particle masses are much more massive than the mass of a star, typically 10$^4-$10$^5$ M$_{\odot}$, once formed, each stellar particle is
 considered as a single stellar population, within which the individual stellar masses are distributed according to the \citet{Miller&Scalo1979} IMF. Stellar 
particles eject metals and thermal energy through stellar winds and Type II and Ia SNe explosions. Each star more massive than 8~M$_{\odot}$ is assumed
 to dump into the ISM, instantaneously, 2$\times$10$^{51}$~erg of thermal energy; 10$^{51}$~erg comes from the stellar wind, and the other 10$^{51}$~erg
 from the SN explosion.
 Moreover, the star is assumed to eject 1.3~M$_{\odot}$ of metals. For the assumed \citet{Miller&Scalo1979} initial mass function, IMF, a stellar particle 
of 10$^5$~M$_{\odot}$ produces 749 Type II SNe. For a more detailed discussion of the processes implemented in the code, see \citet{Kravtsov2003,Kravtsov2005}.
Stellar particles dump energy in the form of heat to the cells in which they are born. If sub-grid physics are not properly simulated, most of this thermal 
energy, inside the cell, is radiated away.  Thus, to allow for outflows, it is common to adopt the strategy of turning off the cooling during a time t$_{off}$
 in the cells where young stellar particles (age~$<$~t$_{off}$) are placed \citep[see][]{Colin2010}. This mechanism along with a relatively high value of
 $\epsilon_{SF}$ allow the gas to expand and move away
 from the star-forming region. As t$_{off}$ can be linked to the crossing time in the cell at the finest grid, 
we could see this parameter as depending on resolution in the sense that the higher the resolution, the smaller its value. In simulations presented
here the cooling is stopped for 40~Myr. Although it has been demonstrated that for spatial resolutions similar to the ones in our models this process 
is unnecessary when simulating dwarf galactic systems \citep{GonzalezSamaniego2014}, a more detailed study is necessary to ensure that the same occurs when simulating MW-sized galaxies.\\

\subsection{Simulation technique and halo selection}

 Simulations presented here have been
 run in a $\Lambda$CDM cosmology with $\Omega_0$ = 0.3, $\Omega_{\Lambda}$ = 0.7, $\Omega_b$= 0.045, and h = 0.7. The CDM power spectrum was taken from
 \citet{Klypin&Holtzman1997} and it was normalized to $\sigma_8$ = 0.8, where $\sigma_8$ is the rms amplitude of mass fluctuations in 8~Mpc~h$^{-1}$ spheres.
To maximize resolution efficiency, we used the zoom-in technique. First, a low-resolution cosmological simulation with only dark matter (DM) particles
 was ran, and then regions  of interest (DM halos) were picked up to be re-simulated with high resolution and with the physics of the gas included. 
The low-resolution simulation was run with 128$^3$ particles inside a box of 20~Mpc~h$^{-1}$ per side, with the box initially covered by a mesh of 128$^3$
cells (zeroth level cells). At z=0, we searched for MW-like mass halos ($7.0\times$10$^{11}$~M$_{\odot}$ $\le$ M$_{vir}$ $\le$ $1.5\times$10$^{12}$~M$_{\odot}$) that 
was not contained within larger halos (distinct halos). We selected only halos that have had not major mergers since z=1.5 and that at z=0 have not a 
a similar mass companion inside a sphere of 1~Mpc~h$^{-1}$. Other restrictions we have imposed are that halos need to be in a
 filament or a wall
not in a void or a knot. After this selection, a Lagrangian region of 3~R$_{vir}$ was identified at z=60 and re-sampled with additional small-scale modes
 \citep{Klypin2002}.
The virial radius in the low resolution runs, R$_{vir}$, is defined as the radius that encloses a mean density equal to $\Delta_{vir}$ times the mean density of the universe,
 where $\Delta_{vir}$ is a quantity that depends on $\Omega_0$, $\Omega_{\Lambda}$, and z. For example, for our cosmology $\Delta_{vir}$(z~=~0)~=~338
 and $\Delta_{vir}$(z~=~1)~=~203. The number of DM particles in the
 high resolution region depends on the number of DM mass species and the mass of the halo. For models with four or five DM mass species this value varies
 from $\sim$ 1.5$\times$10$^6$ (model G.242) to about 7.0$\times$10$^6$ (model G.321). The corresponding
 DM first specie mass per particle of modeled galaxies (m$_{DM1sp}$) is given in Tab.~\ref{tab:1}.\\

 In ART, the initially uniform grid is refined recursively as the matter
 distribution evolves. The cell is refined when the
mass in DM particles exceeds 81.92~(1-F$_{b,U}$)~m$_p$/f$_{DM}$ or when the mass in gas is higher than 81.92~F$_{b,U}$~m$_p$/f$_{g}$, where F$_{b,U}$ is the universal baryon fraction,
 m$_p$=7.75$\times$10$^4$~M$_{\odot}$~h$^{-1}$ is the mass of the lightest particle specie in the DM only simulation, and f$_{DM}$ and f$_{g}$ are factors that control the refinement
aggresivitiy in the dark matter and gas components, respectively. In our models F$_{b,U}$=0.15 according to the cosmology we have imposed and 
the mean DM and gas densities in the box are assumed to be equal to the corresponding universal averages. For the simulations presented in this work the grid is always unconditionally refined to 
the fourth level, corresponding to an effective resolution of 2048$^3$ DM particles. A limit also exists for the maximum allowed
refinement level and it is what will give us the spatial size of the finest grid cells. Here we allow the refinement to go up to
the 11 refinement level what corresponds to a spatial size of the finest grid cells of 109 comoving pc.
Our main model is a re-simulation of a MW-like halo of M$_{vir}~\sim$7.4$\times$10$^{11}$~M$_{\odot}$. However we also re-simulated other 
halos with masses in between 13.9$\times$10$^{11}$~M$_{\odot}$ and 2.0$\times$10$^{11}$~M$_{\odot}$. 
We have used these simulations to find how the general properties of the galactic system change with mass. For each one of these models we have also made simulations changing 
initial parameters such as $\epsilon_{SF}$, n$_{SF}$, 
resolution or number of DM mass species. This set of models helps us to assess how the final results depend on the initial conditions.
 A more detailed study of the effects of changing the simulation parameters can be find in \citet{GonzalezSamaniego2014}. 
Tab.~\ref{tab:1} summarizes the properties of the main MW-sized simulations presented here.\\

Although our MW-sized simulations are realistic, some physical processes are not well implemented in the code we have used. For instance,
 the AGN feedback, the molecular cooling down to 200~K, the formation of molecular H$_2$ clouds, a deep chemical treatment, the
 radiation pressure or the photoionization from massive stars. However, the last two are now implemented in a
 new version of the code and a new set of simulations is being generated. 

In this work we study MW-sized systems, however it is not our goal to reproduce all observed properties.  Here we define a MW-sized system as the one that has a total mass,
 v$_{max}$ and disk with similar mass and scale lengths as the ones observed. Following previous works \citep[e.g. ][]{Robertson2004,Governato2004,Okamoto2005,Scannapieco2009,Guedes2011,Moster2014,Aumer2013,Stinson2013,Vogelsberger2014} we have selected a DM halo that in the high resolution run at z=0 has no other massive halos inside a 800~kpc sphere and that have had a quiescent recent assembling history. We note that our initial conditions do not reflect
the environment in which the MW is embedded but this is certainly not needed
if we just aim at producing a system with a non-negligible disc component.
This latter is almost always achieved under the requirement that the system
evolves in relative isolation in last 10 Gyr or so \citep[e.g.][]{Moster2014,Aumer2013,Stinson2013,Vogelsberger2014}. These
specific systems will be our main MW-sized models. Details of each realization will depend on selected subgrid
 physical parameters, environment and assembling histories.

\begin{figure}
\hspace{-0.5cm}
\includegraphics[scale=0.3]{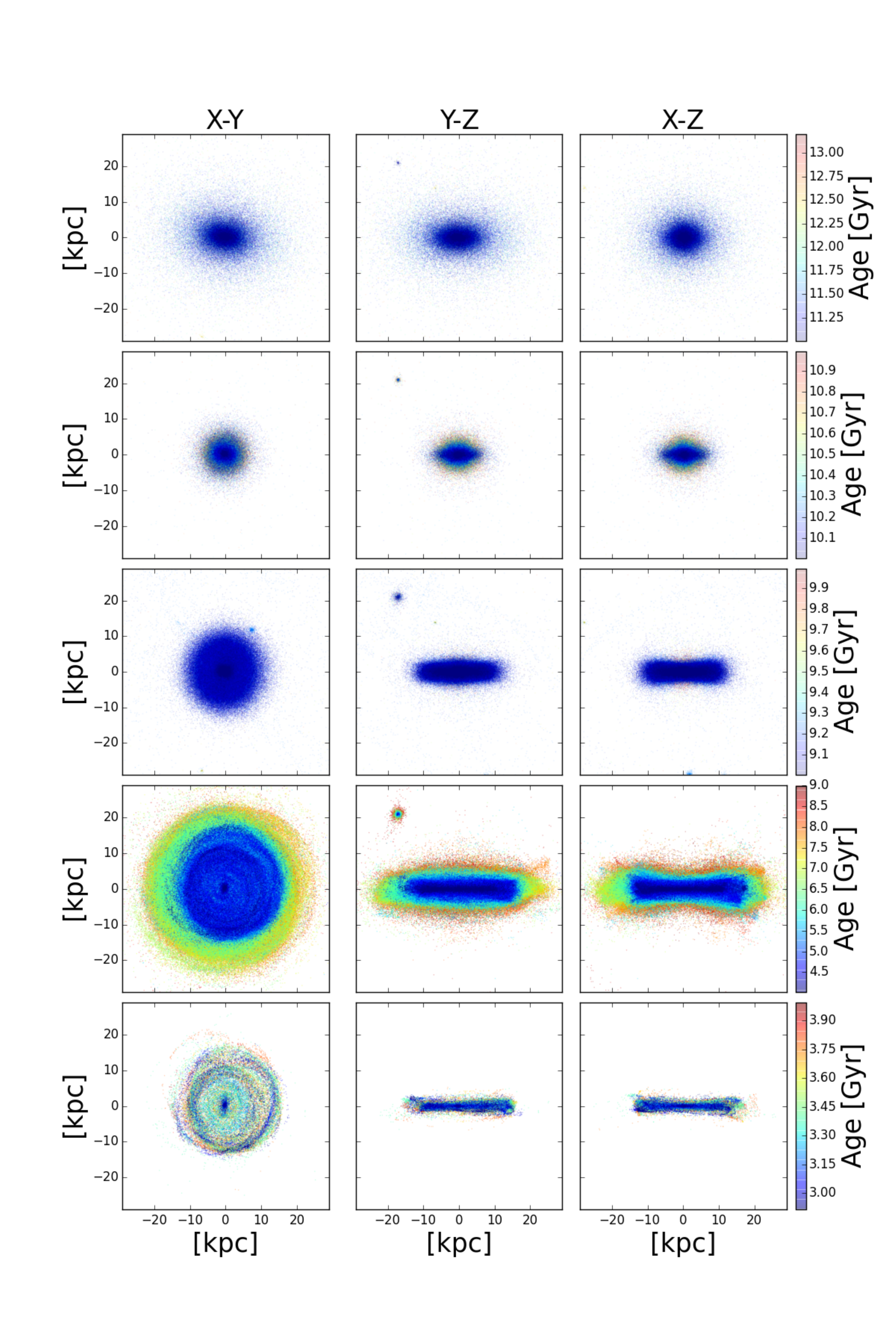}\\
\caption{Face on (first column) and edge on view (last two columns) of stars in our simulated Milky Way sized galaxy. Each row correspond to a different stellar age. From top to the bottom,  
11$-$13.467~Gyr, 10$-$11~Gyr, 9$-$10~Gyr, 4$-$9~Gyr and 0$-$4~Gyr.
 Color code indicates stellar age.}
\label{fig:1}
\end{figure}

\begin{figure}
\centering
\includegraphics[scale=0.7]{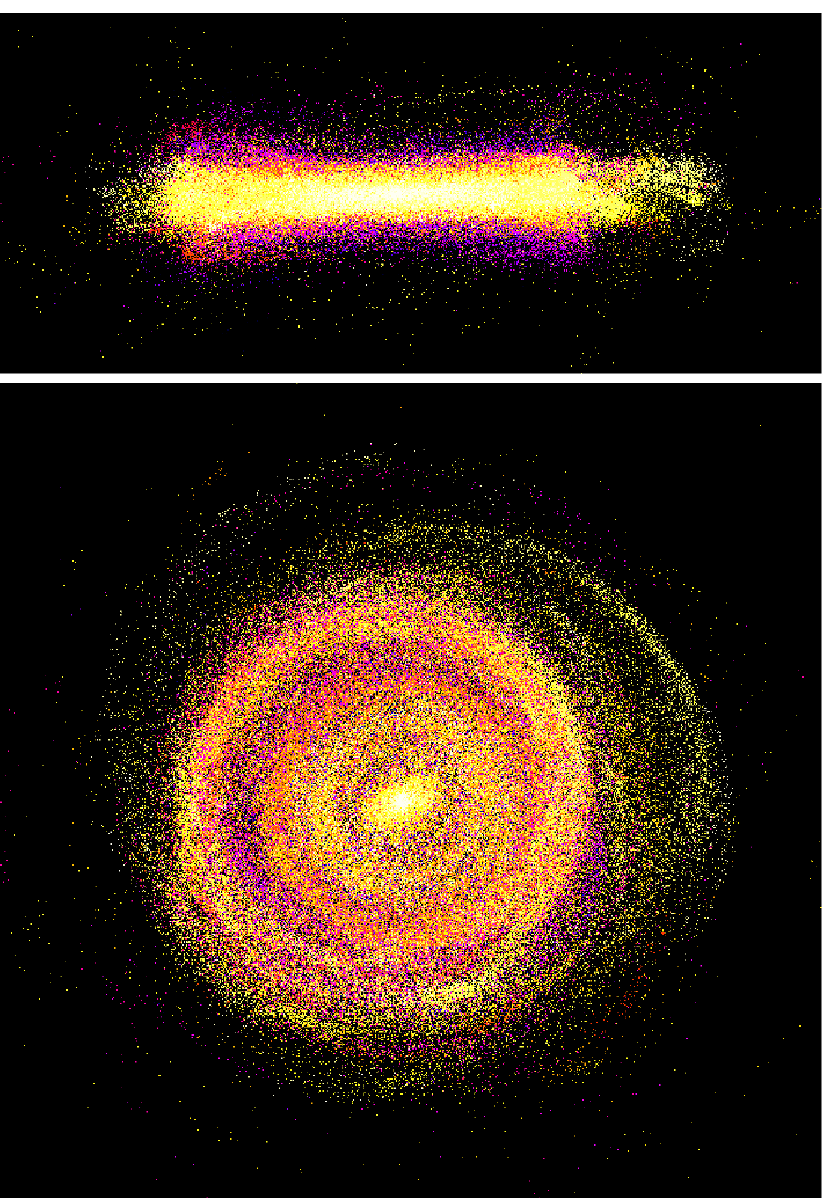}
\caption{Edge on (top) and face on (bottom) views of a young stellar population (0$-$7~Gyr) of model G.323 at z=0. 
 Both Panels span 50~kpc in the x-axis. The y-axis span 22~kpc in the top panel and 50~kpc in the bottom. Color code indicates stellar 
age, qualitatively.}
\label{fig:2}
\end{figure}

\begin{figure}
\hspace{-0.5cm}
\includegraphics[scale=0.4]{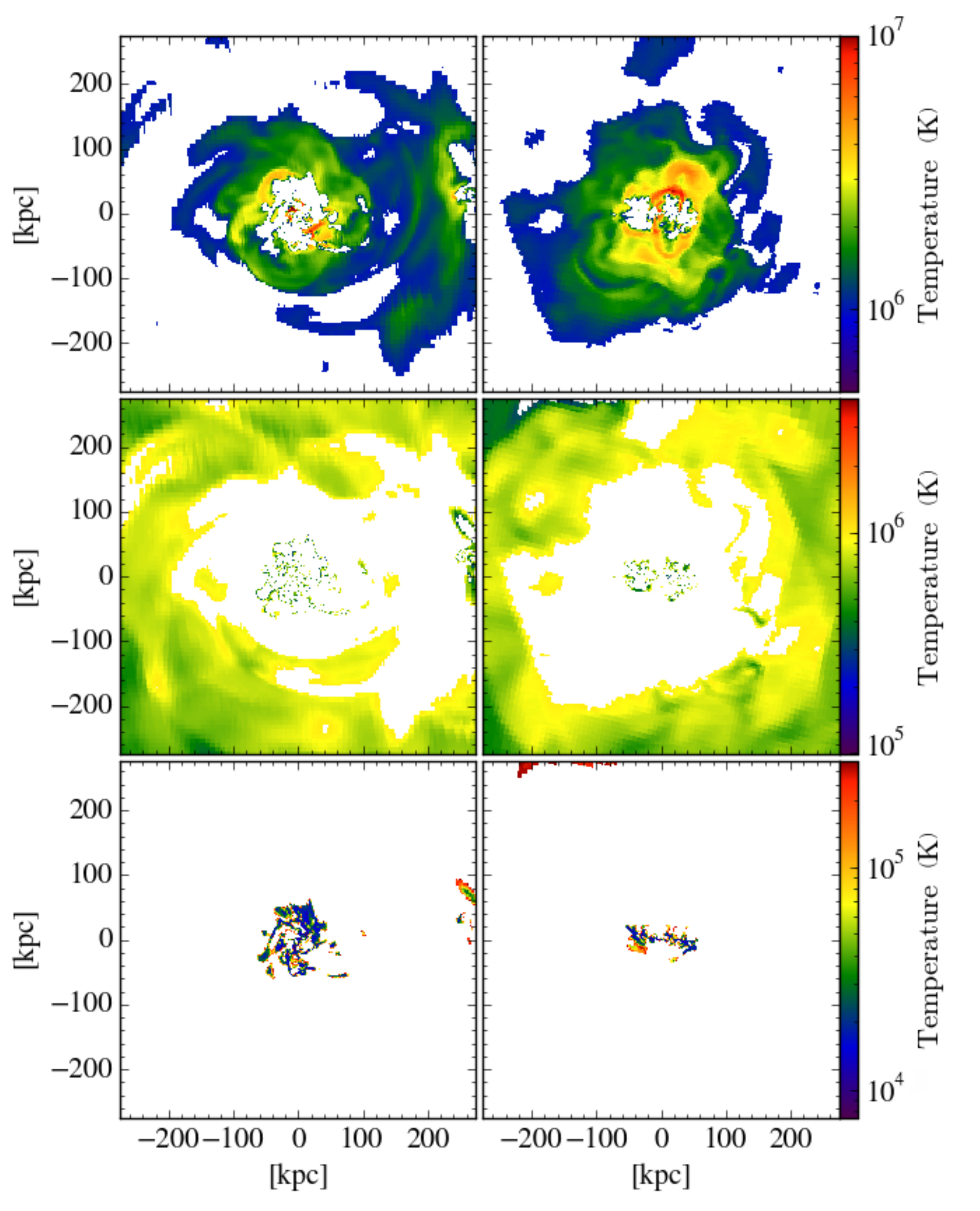}
\caption{Face on (left) and edge on (right) views of gas density at z=0 in model G.323, as function of its temperature. From top to the bottom,
 10$^6-$10$^7$~K, 3$\times$10$^5-$10$^6$~K and 10$^3-$3$\times$10$^5$~K.  In color we show the gas temperature.}
\label{fig:6}
\end{figure}

\section{GARROTXA MW-sized runs}\label{sec:results}

\subsection{General properties}\label{sec:genprop}

In this work we focus mostly in three MW-sized models (G.3 series) that differ only in the aggresivity on the refinement criteria. Parameters that
 control the refinement aggresivity (f$_{DM}$, f$_g$) take the following values: (1.,1.) for the less aggressive model, (4.,8.) 
for the intermediate and (8., 32.) for the more aggressive. These models are the ones we have found 
to be in a better agreement with observed properties of the MW. The spatial resolution of all our models is high, around 109~pc.
 Model G.321 is the one with a lightest refinement, G.322 has intermediate
 conditions and G.323 is the one with hardest refinement conditions (harder means that the code uses a lower mass threshold for the refinement).
 All parameters we present in this section are summarized and compared
 with recent MW-sized models and with its observational values in Tab.~\ref{tab:1}.  In the table we see that there are parameters
 which are very sensitive to a change in the refinement setting as, for example, the SFR(z=0) that changes from 0.02  to 
0.32, G.322 and G.323, respectively. Yet, most parameters do converge as can be seen in Tab.~\ref{tab:1}; in particular, 
circular velocities profiles agree within 5\%  (see Appendix A for more information).\\
 To ensure numerical convergence we also generated a low resolution model. Comparison between models can be found in the Appendix A.\\

\subsubsection{Morphology}\label{sec:morfologia}

At z$\sim$0 our three G.3 models are massive spiral galaxies with several non-axisymmetric structures in the disk such as bars or spirals (see Fig.~\ref{fig:1}
 and \ref{fig:2}). 
Their assembling history is quiet after z=1.5 and the last major merger occurs at z=3. In Fig.~\ref{fig:1} we show the face on (first column) and edge on 
(last two columns) views of stars of our model G.323 at z=0. For the other two models the main picture is similar but with a smaller number of stellar particles. 
Each row correspond to a different stellar age (see figure caption). In this figure it can be observed that
 young stellar component (0$-$4~Gyr) is distributed in a flat disk structure. On the other hand, older stellar populations (4$-$10~Gyr)
are also distributed into a disk structure but in these cases not as flat as the younger. We argue that disk scale height and length
 depends on the age of the stellar population and that the former is higher and the latter is smaller when the population gets older,
 this issue will be addressed in Sec.~\ref{sec:6.2.3.2}. Another interesting property that can be easily observed 
in the edge on views of our models is that stellar population has an evident flare that is more evident for older populations. The face on view of 
the younger stellar population shows the presence of rings, spiral arms and also a young bar, in the disk. Spirals, rings and the young bar can be better observed in models G.322 and G.323 where the number of stellar
particles is higher (see Fig.~\ref{fig:2}). The young bar has grown from disk particles that have ages between 0 and 8$-$9~Gyr. The old stellar population (11$-$13.476~Gyr) is distributed in a bar/ellipsoid component and 
it has been originated in the last major merger at z=3 (Valenzuela et al. in preparation). In Fig.~\ref{fig:6} we show how the gas component as function of temperature (see figure caption). In this figure is easy to see that gas distribution is not isotropic. Gas is distributed in different regions of the system depending on its temperature: cold gas is present
 in the young stellar disk region and hot gas fills the out-of-plane region and is embedded in the DM halo. It is also interesting to see how cold/intermediate gas has a warped structure that is 
also marginally observed in the stellar component.

\subsubsection{Stellar, gas and dark matter virial mass}

Here we define the virial radius (r$_{vir}$) as the one where the sphere of radius r$_{vir}$ encloses a mean density 97 times denser than the critical density 
 ($\rho_{crit}$) of a spatially flat Universe $\rho_{crit}$=3H$^2$(z)/(8$\pi$G). We have used value 97 as it is the value derived from the spherical top-hat collapse model for $\Lambda$CDM at 
z=0 for our cosmology \citep{Bryan&Norman1998}. This virial radius definition is borrowed from structure growth theory and then its use is not appropriated when one
 wants to define a physically meaningful halo edge \citep{Cuesta2008,Zemp2013}. To avoid this problem we also give the properties for another commonly used 
definition which is r$_{200}$, the radius that encloses a mean density equal to 200 times $\rho_{crit}$. Using the first definition we have obtained that the
 virial radius at z=0 is r$_{vir}\sim$230~kpc in our three main models and that the mass enclosed in this radius is 
M$_{vir}$=7.20$-$7.61$\times$10$^{11}$~M$_{\odot}$. This value for the M$_{vir}$ falls well inside the observational range that is M$_{vir}$=0.6$-$2.4$\times$10$^{12}$~M$_{\odot}$ by \citet{Xue2008}, 
\citet{Kafle2012}, \citet{Boylan-Kolchin2013} and \citet{Kafle2014}. The total mass enclosed at 
r$_{200}$=175.6~kpc is M$_{200}$=6.71$-$6.90$\times$10$^{11}$~M$_{\odot}$. Total virial mass is distributed in dark, stellar and gaseous matter as follows: M$_{DM}$=5.86$-$6.79$\times$10$^{11}$~M$_{\odot}$,
M$_{*}$=6.1$-$6.2$\times$10$^{10}$~M$_{\odot}$ and M$_{gas}$=1.73$-$2.70$\times$10$^{10}$~M$_{\odot}$. The baryonic fraction of our G.3 models is F$_{b,U}$=0.104$-$0.121. These values are 19$-$31\% smaller than the universal
value for the adopted cosmology which is F$_{b,U}$=0.15. Inside r$_{vir}$, and for the model G.321, we have a total number of particles of 
N$_{total}$=7.52$\times$10$^{6}$, where N$_{DM1sp}$=7.12$\times$10$^{6}$ and  N$_{*}$=3.94$\times$10$^{5}$ and a total number of gas cells of 
2.0$\times$10$^{6}$. For models with a more aggressive refinement the number of stellar particles and gas cells grow up to 2.34$\times$10$^{6}$ and 
6.8$\times$10$^{6}$, respectively. All dark matter particles inside virial radius
belong to less massive DM specie (1sp) and have a mass of 9.25$\times$10$^{4}$~M$_{\odot}$. Star particles have masses between $\sim$10$^{3}$ and 
1.2$\times$10$^{6}$~M$_{\odot}$. All these parameters are summarized in Tab.~\ref{tab:1}.

\subsubsection{Circular velocity curves}

\begin{figure*}
\centering
\includegraphics[scale=0.33]{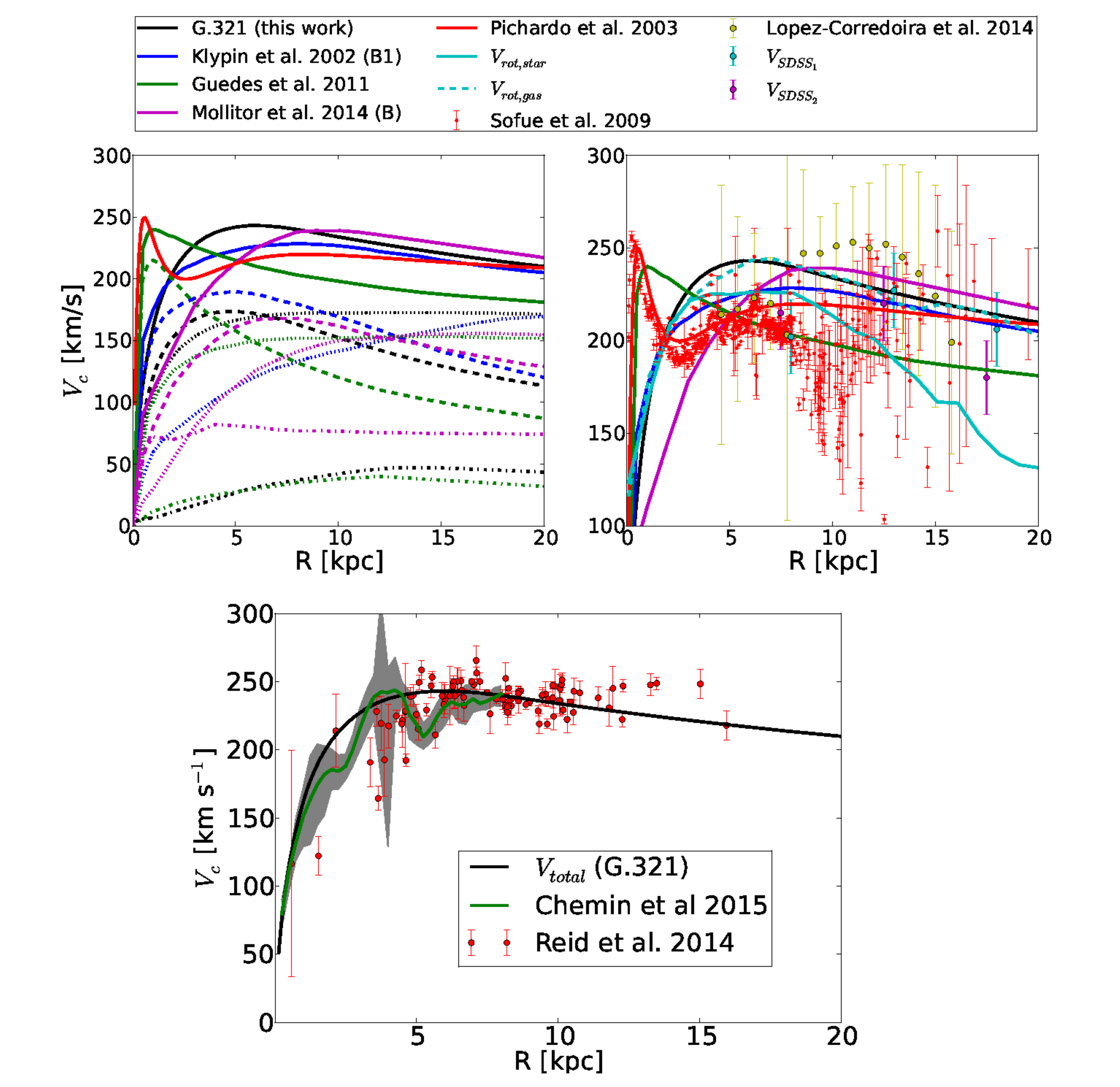}
\caption{Top-left: Circular velocity curve of our simulated Milky Way sized galaxy model G.321 (black), \citet{Guedes2011} (green), model B1 in \citet{Klypin2002} (blue), \citet{Mollitor2014} 
(magenta) and analytical model in \citet{Pichardo2003} (red). The figure shows the contribution to the circular velocity V$_c$=$\sqrt{GM_{<r}/r}$ 
 of dark matter (short dashed curve),
stars (long dashed), gas (dot-dashed) and total (solid curve) mass components. Top-right: Total circular velocity curves of the same models shown at left, observational 
data points and gas and stars rotation curves of our model G.321 (cyan dashed and solid lines, respectively). Data points come
 from two realizations of the rotation curve of the Milky Way from observations of blue horizontal-branch halo stars in the SDSS \citep{Xue2008}, in cyan and magenta
dots, from \citet{LopezCorredoira2014}, in green, and from \citet{Sofue2009}, in red. Bottom: Observational circular velocity curve obtained using massive young 
star forming regions of the MW \citep{Reid2014}. The total V$_c$ of our model
G.321 at z=0 is shown as a black solid line. In green we show the hypothetical gaseous rotation curve of the Milky Way obtained
 by \citet{Chemin2015} after correcting the non-circular motions in the velocity profile of the Milky Way inferred with the
tangent-point method by \citet{Kalberla2005}. In grey we show \citet{Chemin2015} curve 1$\sigma$ range.}
\label{fig:4}
\end{figure*}

Circular velocity curves of our main simulations, computed using the $\sqrt{GM_{<r}/r}$ approach, are shown in Fig.~\ref{fig:4} top left panel, as a solid black line. 
Circular velocity curves have also 
been computed using the real galactic potential by using Tipsy package. Results obtained using both techniques are in good agreement.
Fig.~\ref{fig:4} also shows circular velocity 
curves from other state of the art MW-sized simulations: \citet{Klypin2002} B1 model (blue), \citet{Mollitor2014} model (magenta), 
\citet{Guedes2011} ERIS
simulation (green). As a comparison we show in red the analytical fit to the \citet{Sofue2009} MW rotation curve data, presented in 
\citet{Pichardo2003}. Solid lines are the total circular velocity curves while the DM, stellar and gas 
components are shown as long dashed, dotted and short dashed lines, respectively. In the top right panel we show, first, G.321 
velocity rotation curves of stars and gas, computed as the mean of the tangential velocity in rings centered to the galactic 
center (cyan solid and dashed lines), and the 
total circular velocity curve obtained using V$_c$=$\sqrt{GM_{<r}/r}$ approach. As expected, gas follows the circular velocity 
while stars are affected by the asymmetric drift. In this panel we also compare stellar rotation curve of our simulation
(cyan solid line) with values from MW observations. We show rotation velocity curve estimations from blue
 horizontal-branch halo stars in the SDSS \citep{Xue2008} (cyan and magenta dots), from \citet{Sofue2009} (red dots) and 
from observations of \citet{LopezCorredoira2014} (green dots). In the bottom panel we compare the recent data
for the circular velocity curve obtained by \citet{Reid2014}, using high mass star forming regions, with the total circular velocity curve of our G.321 model.
 As can be seen in the figure our G.3 models are in a a very good agreement with the most recent observational data. We also show the recent
hypothetical gaseous rotation curve of the Milky Way obtained by \citet{Chemin2015} after correcting for bar non-circular motions.
 It is important to mention that we only show results for G.321 model because circular velocity curves of models
 G.322 and G.323 do not differ significantly.
  The peak of z=0 circular velocity of our models is reached at R$_{peak}\sim$5.69~kpc with a value of V$_c$(R$_{peak}$)=237.5$-$243.8~km~s$^{-1}$, the value at a standard solar 
radius (R$_{\odot}$=8~kpc) is V$_{c\odot}$=233.3$-$239.8~km~s$^{-1}$. Also the ratio between circular velocity at 2.2 times disk scale radius (V$_{2.2}$) and circular velocity at r$_{200}$ (V$_{200}$)
of our simulations (V$_{2.2}$/V$_{200}\sim$1.9) is inside the observational range for the MW that is 1.67$^{+0.31}_ {-0.24}$~$-$~1.11$^{+0.22}_ {-0.20}$ \citep{Xue2008,Dutton2010}.\\
An interesting exercise has been to compare circular velocity curves of our simulations with the one of  \citet{Guedes2011} model.
We have easily seen that our simulations have a bit higher velocity curves and do not present a peak in the inner regions as in
 \citet{Guedes2011}. In simulations this internal peak is usually associated with the presence of a massive bulge in the central
parts of the simulated galactic system. On the other hand, a similar peak in the V$_c$ is shown in observational work of 
\citet{Sofue2009}. It is in discussion whereas this peak detected in \citet{Sofue2009} observations 
is a signature of non-circular motions inside the bar or of a massive bulge \citep{Duval1983,Chemin2015}. Non-circular motions need to
 be computed to solve such question. We plan to undertake this work in a near future following the one started by \citet{Chemin2015}.\\
Finally, despite of the rotation and circular velocity curves we introduce here fall well inside observational ranges it is 
important to take into account that nowadays observational uncertainties are still high.
  
\newpage
\clearpage
\LTcapwidth=\textwidth
\begin{scriptsize}
\begin{longtable*}{lccccccc}
\\
\hline
\hline
                        &r$_{vir}$&  M$_{vir}$                  & M$_{*}$                 & M$_{gas}$           & M$_{hotgas}$       & M$_{warmgas}$ &M$_{coldgas}$     \\
                        & [kpc]   & [M$_{\odot}$]               & [M$_{\odot}$]           & [M$_{\odot}$]       & [M$_{\odot}$]      & [M$_{\odot}$]   &[M$_{\odot}$]         \\
\hline
 &          &            &             &                    &            &                         &         \\
G.321               & 230.1  & 7.33$\times$10$^{11}$        & 6.1$\times$10$^{10}$     & 2.70$\times$10$^{10}$ & 1.22$\times$10$^{10}$&  5.66$\times$10$^{9}$ &9.34$\times$10$^{9}$   \\
G.322               & 230.1  & 7.20$\times$10$^{11}$        & 6.1$\times$10$^{10}$     & 1.73$\times$10$^{10}$& 0.98$\times$10$^{10}$ &  5.98$\times$10$^{9}$ &1.52$\times$10$^{9}$    \\
G.323               & 230.1  & 7.61$\times$10$^{11}$        & 6.2$\times$10$^{10}$     & 1.96$\times$10$^{10}$ & 1.32$\times$10$^{10}$& 4.54$\times$10$^{9}$  &1.86$\times$10$^{9}$   \\
ERIS & 239.0   & 7.90$\times$10$^{11}$         & 3.9$\times$10$^{10}$     & 5.69$\times$10$^{10}$& 3.60$\times$10$^{10}$& 14.20$\times$10$^{9}$ & 6.70$\times$10$^{9}$  \\
MollitorB& 234.0   & 7.10$\times$10$^{11}$         & 5.6$\times$10$^{10}$     & 7.96$\times$10$^{10}$&   $-$                 & $-$&          $-$                 \\                                                                                                                                                         
MW obs.            &  $-$       & 1.0$\pm$0.30$\times$10$^{12}$ & 4.9-5.5$\times$10$^{10}$ &         $-$            &      $-$              & $-$  &              7.3$-$9.5$\times$10$^{9}$  \\  
\hline
\hline
                        & r$_{200}$ & M$_{200}$        & F$_{b,U}$ & m$_{DM1sp}$         & m$_{*min}$          & m$_{*max}$          & m$_{SPH}$        \\
                        & [kpc]     & [M$_{\odot}$]    &         & [M$_{\odot}$]       & [M$_{\odot}$]       & [M$_{\odot}$]       & [M$_{\odot}$] \\

\hline
 &          &            &             &                    &            &                         &         \\
G.321                 & 175.6 & 6.84$\times$10$^{11}$& 0.120   & 0.93$\times$10$^{5}$& 0.24$\times$10$^{4}$& 120.0$\times$10$^{4}$& $-$              \\
G.322                 & 175.6 & 6.71$\times$10$^{11}$& 0.109   & 0.93$\times$10$^{5}$& 0.12$\times$10$^{4}$& 120.0$\times$10$^{4}$& $-$               \\
G.323                & 175.6 & 6.90$\times$10$^{11}$& 0.107   & 0.93$\times$10$^{5}$& 0.12$\times$10$^{4}$& 120.0$\times$10$^{4}$& $-$                \\
ERIS& 175.0 &   6.60$\times$10$^{11}$ & 0.120   & 0.98$\times$10$^{5}$ & 0.63$\times$10$^{4}$ & 0.63$\times$10$^{4}$& 2.0$\times$10$^{4}$\\
MollitorB& 176.5 & 6.28$\times$10$^{11}$& 0.191        & 2.30$\times$10$^{5}$ & 4.50$\times$10$^{4}$ &  4.50$\times$10$^{4}$& $-$                  \\                                                                                                                                                         
MW obs.  &   $-$      & $-$ & $-$ &        $-$             &       $-$             &  $-$ &        $-$                         \\
\hline
\hline
                        & N$_{part}$         & N$_{DM1sp}$        & N$_{*}$            & N$_{gas}$                &  Resolution     &  CPU time        &  CODE      \\
                        & [10$^{6}$]         &  [10$^{6}$]        & [10$^{6}$]         & [10$^{6}$]               & [pc]            &  [10$^4$ h]  &               \\
\hline
 &          &            &             &                    &            &                                        &         \\
G.321               & 7.52                 & 7.12               & 0.39               &  2.0                     & 109(1cell)     & 2.5             &ART + hydro     \\
G.322               & 7.95                 & 7.08               & 0.86               &  5.0                     & 109(1cell)     & 4.3             &ART + hydro     \\
G.323               & 9.44                  & 7.10               & 2.34               &  6.80                    & 109(1cell)     & 9.2           &ART + hydro     \\
ERIS & 15.60 (+SPH)      & 7.00               & 8.60               & 3.0                      &120($\epsilon$)&160.0             &GASOLINE (SPH)    \\
MollitorB& 6.06                & 3.91              &  2.15               & 2.50                      & 150(1cell)     &   $-$              & RAMSES (hydro)    \\                                                                                                                                                         
MW obs.    &      $-$    &        $-$    &     $\sim$100000.0        &        $-$            &     $-$       &           $-$              &   $-$      \\

\hline
\hline
                          & c=r$_{vir}$/r$_s$ & R$_d$            & h$_{z,young}$    & h$_{z,old}$    &  M$_{hotgas}$/M$_{vir}$ & $\alpha_X$ & SFR (z=0)\\ 
                          &                   & [kpc]            & [pc]            & [pc]             &                    &                    & [M$_{\odot}$ yr$^{-1}$] \\
\hline 
 &          &            &             &                    &            &                         &         \\
G.321                   & 28.5              & 2.56 (4.89/2.21) & 277 (exp)            & 1356       &       0.016        & -0.62              & 0.27 \\
G.322                   & 26.8              & 3.20 (5.26/2.26) & 295 (exp)           & 960        &      0.013           & -0.83              & 0.02\\
G.323                   & 26.9              & 3.03 (3.19/3.31) & 393 (exp)           & 1048        &      0.017           & -0.67              & 0.32 \\
ERIS  & 22.0              & 2.50                 & 490 (sech$^2$)       &      $-$  &       0.017          &  -1.13             & 1.10 \\
MollitorB  & 56.5              & 3.39                  &     $-$            &        $-$  &       0.046           &   -4.54                 & 4.75 \\
MW obs.    &  21.1$^{+14.8}_{-8.3}$   & 2.0$-$4.5  & 300$\pm$60       & 600$-$1100$\pm$60&       $-$           &     $-$            & 0.68-1.45 \\
\hline
\hline
                          &  V$_{c\odot}$(R=8~kpc) & R$_{peak}$ & V$_{c}$(R$_{peak}$) & V$_{2.2}$/V$_{200}$                                    &   n$_{SF}$    & T$_{SF}$  & $\epsilon_{SF}$ \\
                          & [km s$^{-1}$]         &  [kpc]      & [km s$^{-1}$]       &                                                        &    [cm$^{-3}$] & [10$^3$ K]&                  \\
\hline 
 &          &            &             &                    &            &                         &         \\
G.321                   & 239.8                 &  5.69       & 243.8               & 1.90                                                   &      1.0           & 9         & 0.65            \\
G.322                   &  233.6               &  5.69       & 237.5               & 1.85                                                   &     1.0           & 9         & 0.65            \\
G.323                   & 233.3                 &  5.69      & 237.8               & 1.93                                                   &     1.0           & 9         & 0.65            \\
ERIS  &  206.0                 & 1.34       & 238.0               & 1.66                                                   &      5.0           & 30        & 0.10             \\
MollitorB  &  233.0                 &    9.63       & 233.0               &        $-$                                           &    2.7         & 3         & 0.01             \\    
MW obs.          &  221$\pm$18            &    $-$          &        $-$             & 1.11$^{+0.22}_ {-0.20}$&    $-$      &     $-$       &       $-$            \\ 
\hline
\hline
                        &   Box    &  z$_{ini}$ &  $\Omega_0$ & $\Omega_{\Lambda}$ & $\Omega_b$ & H$_0$                   & $\sigma_8$\\
                        &  [Mpc h$^{-1}$]&            &              &                    &            & [km~s$^{-1}$~Mpc$^{-1}$]&                  \\
\hline
 &          &            &             &                    &            &                         &         \\
G.321                 &   20     & 60         &   0.30        & 0.70                & 0.045      & 70                      & 0.80  \\
G.322                 &   20     & 60         &  0.30         & 0.70                & 0.045      & 70                      & 0.80               \\
G.323          &   20     & 60         &   0.30        &  0.70               & 0.045      & 70                      & 0.80             \\
ERIS &   90     & 90           &   0.24       & 0.76               & 0.042      & 73                      & 0.76 \\
MollitorB&   20     & 50         &  0.276      & 0.724              & 0.045      & 70.3                    &    $-$       \\
MW obs.  &    $-$      &     $-$       &     $-$        &          $-$          &     $-$       &        $-$                 &   $-$      \\
\hline
\hline
\\
\caption{Parameters of GARROTXA321 (G.321), GARROTXA322 (G.322) and GARROTXA323 (G.323) simulations at z=0.
 r$_{vir}$ is assumed to be equal to r$_{97}$. 
Mass related parameters: M$_{vir}$ is total mass inside r$_{vir}$. M$_{*}$ and M$_{gas}$ are total stellar and gas mass inside r$_{vir}$. Hot, warm and cold gas mass (M$_{hotgas}$ , M$_{warmgas}$, M$_{coldgas}$) are gas mass
 inside r$_{vir}$ with temperature T $>$ 3$\times$10$^5$ K, 
3$\times$10$^5$ K $>$ T $>$ 3$\times$10$^4$ K and T $<$ 3$\times$10$^4$ K, respectively. F$_{b,U}$ is the baryonic fraction (M$_{gas+*}$/M$_{vir}$), m$_{DM1sp}$
is the mass of a single particle belonging to the first DM mass specie, m$_{* min/max}$ is the lowest and highest mass of stellar 
particles and m$_{SPH}$ is the mass of SPH particles. 
Structural parameters: c is the DM halo concentration parameter defined as r$_{vir}$/r$_s$, R$_d$ is the 
disk scale length, h$_{z,young/old}$ is the disk scale height of young (0$-$0.5~Gyr) and old (4.0$-$11.0~Gyr) stellar populations and 
$\alpha_X$ is the hot gas density power-law exponent ($\rho_{hotgas}$(r) $\propto$ r$^{\alpha_X}$). 
Kinematical parameters: V$_{c\odot}$ is the circular velocity at solar position (R$_{\odot}$~=~8~kpc), R$_{peak}$ is the radius in where
circular velocity curve reaches its highest value (V$_{c}$(R$_{peak}$)).
Initial condition parameters: n$_{SF}$, T$_{SF}$ and $\epsilon_{SF}$ are the star formation density and temperature thresholds and the 
star formation efficiency. Box is the simulated box size in Mpc~h$^{-1}$, z$_{ini}$ is the initial redshift and $\Omega_0$, $\Omega_{\Lambda}$, $\Omega_b$, H$_0$ and $\sigma_8$
define the assumed cosmological model.
Comparison with other state of the art simulations is also shown (ERIS simulation \citep{Guedes2011} and model B in \citet{Mollitor2014}).
Observed values for the MW have been obtained from \citet{Ferriere2001,Flynn2006,Du2006,Hammer2007,Xue2008,Juric2008,Dutton2010} and 
\citet{Kafle2014}, see text for more details on references.} \\
\label{tab:1}
\end{longtable*}
\end{scriptsize}


\subsection{Dark matter component}\label{sec:6.2.2}

\begin{figure}
\centering
\includegraphics[scale=0.4]{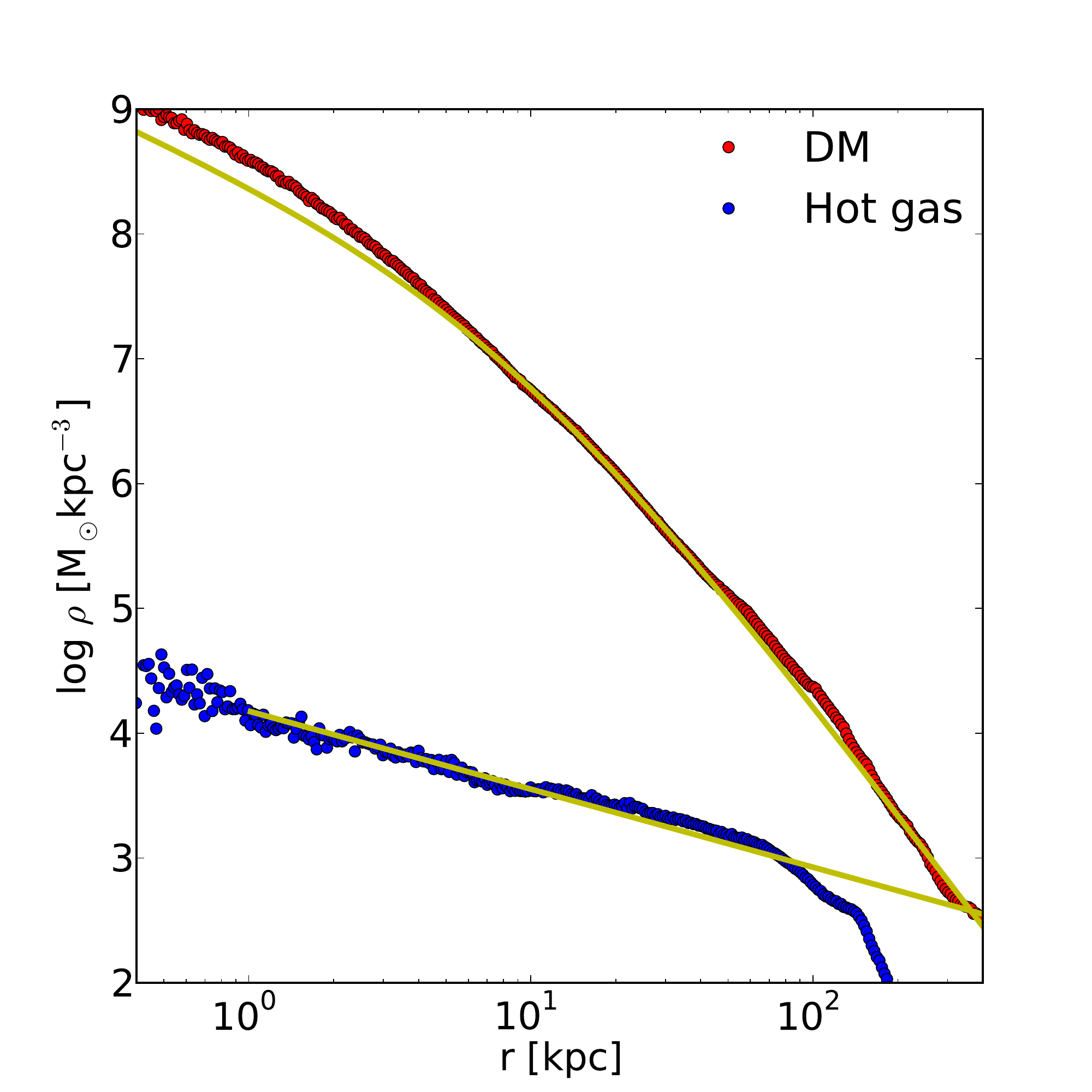}
\caption{Average dark matter (red dots) and hot gas (T $>$ 3$\times$10$^5$ K, blue dots) density profiles at z=0. The upper solid yellow
line show the best NFW profile fit to the dark matter mass distribution, in the range from 5~kpc, outside bar-bulge region, to r$_{vir}$. The lower solid yellow 
line show the best fit power-law profile to the hot gas mass distribution between 5 and 20~kpc. The DM best fit NFW profile is characterized by a large halo concentration parameter 
c=28.5 as the dark matter halo contracts in response to the condensation of baryons in its center (adiabatic contraction). In the inner region it can be 
clearly observed that no DM core is present. The hot gas best fit power-law profile gives a slope of $\alpha_X$=-0.62.}
\label{fig:5}
\end{figure}

As we have mentioned in the last section all dark matter particles inside virial radius are particles of the first DM specie, the less massive one. In Fig.~\ref{fig:5} 
we show the dark matter density profile as a function of radius (red circles) and the best fit of NFW profile for the model G.321 (upper
yellow solid line). We avoid the central region (R~$<$~5~kpc)
 in the fit of the NFW profile to avoid perturbations derived from the presence of baryons (adiabatic contraction).
The best NFW fit gives a large value for the halo concentration that is c=r$_{vir}$/r$_{s}$=28.5, 26.8 and 26.9 for G.321, G.322 and G.323, respectively.
These values fall inside the observational range presented in \citet{Kafle2014} that is 21.1$^{+14.8}_{-8.3}$. No core is observed in the central region of our G.3 models.
The halo spin parameter as defined in \citet{Bullock2001} is $\lambda$'=0.019, 0.017 and 0.022. These $\lambda$' values also fall 
near the range obtained by \citet{Bullock2001} after analyzing more than 500 simulated halos (see figure~1 and 2 in \citet{Bullock2001}).

\subsection{Gas component}\label{sec:6.2.4}

In our models G.321, G.322 and G.323 the total gas mass inside the virial radius is M$_{gas}$=2.7, 1.73 and 1.96 $\times$10$^{10}$~M$_{\odot}$. 
In G.321, 9.34$\times$10$^9$~M$_{\odot}$ of gas mass
is in the cold gas phase (T~$<$~3$\times$10$^4$~K). In G.322 and G.323 cold gas mass is 1.52 and 1.86$\times$10$^9$~M$_{\odot}$. The 
cold gas mass in G.321 is
 comparable to the total mass of the molecular, atomic and warm ionized medium inferred for the MW that is 
$\sim$7.3$-$9.5$\times$10$^9$ M$_{\odot}$ \citep{Ferriere2001}. For the rest of models presented here the amount of cold gas is smaller than
the observed value.
This z=0 cold gas mass reduction is consequence of an increase on the star formation at early ages, that in its turn is provoked by 
the higher aggresivity in the refinement criteria. This increase on the early star formation is the only significant effect we have 
observed when changing the refinement criteria
and it is consequence of resolving a larger number of small dense regions in where the SF criteria is accomplished.\\
At z=0, the SF criteria is satisfied only in the inner disk regions (R~$<$~7~kpc). This spatially limited star formation 
accords with the young stellar distribution shown in Fig.~\ref{fig:1} and also with the decrease of the Star Formation Rate
 (SFR) at z=0 that can be seen in Fig.~\ref{fig:13}.

 The hot gas mass (T~$>$~3$\times$10$^5$~K) is around 1.22, 0.98 and 1.32$\times$10$^{10}$~M$_{\odot}$ for models G.321,
 G.322 and G.323, respectively.  The definition of hot gas we use in this work is observationally motivated: total
 hot gas mass has to be inferred from ionized oxygen observations \citep{Gupta2012} and this is only possible when gas has
opacity in the X-ray region (T~$>$~3$\times$10$^5$~K).
In Fig.~\ref{fig:6} we show the gas distribution
as function of temperature in model G.323. Each panel show the distribution of gas at different temperatures (see figure caption). As can be seen in this figure cold gas is placed in disk region while warm-hot is embedded within the dark matter halo. Contrary to the standard assumption, hot gas
 do not follow the dark matter radial distribution. This can be seen in Fig.~\ref{fig:5} where we show the dark matter density distribution (red) and the hot 
gas density distribution (blue). The hot gas density distribution inside r~=~100~kpc can be fitted by a power
law ($\rho_{hotgas}$(r)~$\propto$~r$^{\alpha_X}$). After fitting data from models G.321, G.322 and G.323 we have obtained that hot gas density distribution scale factors ($\alpha_X$) of these models are
 -0.62, -0.83 and -0.67, respectively. These results are slightly higher than values used to fit analytical models of hot gas in dark matter halos \citep[-0.9 in][]{Anderson&Bregman2010} and the ones 
obtained in other similar models like ERIS 
\citep[-1.1 in ][]{Guedes2011}. The hot gas density distribution power law fit for model G.321 is shown in Fig.~\ref{fig:5} (bottom yellow straight line).
 
In Sec.~\ref{sec:results1} we present our first results on the study of hot gas distribution and a comparison with observational values.

\subsection{Stellar component}\label{sec:6.2.3}

\begin{figure}
\centering
\includegraphics[scale=0.3]{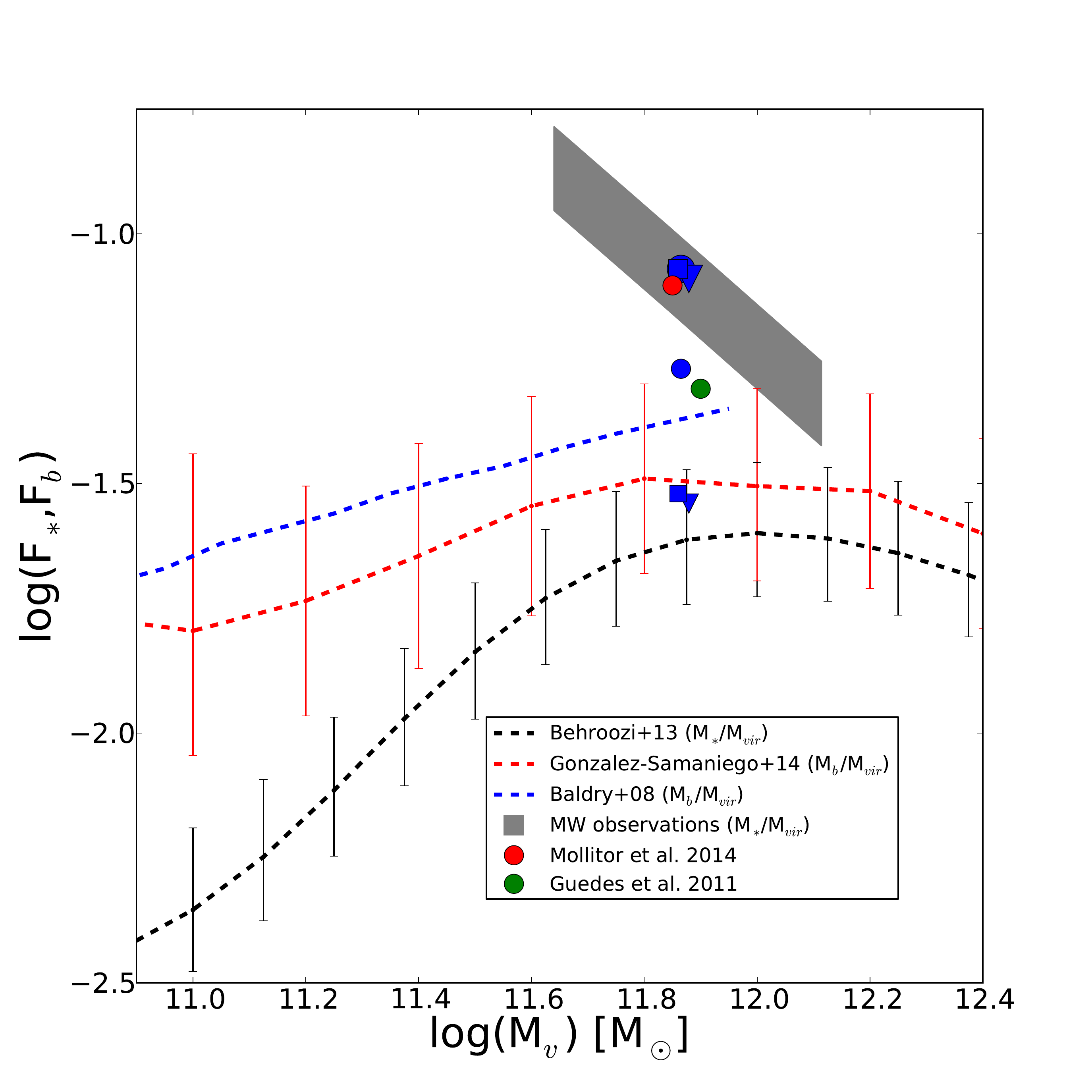}
\caption{Comparison between stellar/baryonic fraction (F$_{*,b}$~=~M$_{*,b}$/M$_{vir}$) theoretical predictions and the ones obtained from observations and simulations. 
Blue items show M$_*$/M$_{vir}$ values in our models G.321 (circles), G.322 (triangles) and  G.323 (squares), using total stellar mass (large items)
and only disk stellar mass (small items), see Sec.~\ref{sec:6.2.3.1} for information about stellar disk selection process. Observed range for the M$_*$/M$_{vir}$ in the Milky Way Galaxy \citep{Flynn2006,Xue2008,Liquia2014,Gibbons2014} is shown as a grey shadowed 
region. Red dot show the M$_*$/M$_{vir}$ value obtained in model B presented by \citet{Mollitor2014}. Green dot shows M$_*$/M$_{vir}$ of the ERIS simulation
 \citep{Guedes2011}. \citet{Behroozi2013} observational M$_*$/M$_{vir}$ curve inferred from observations at z=0.1, with no distinction between
blue or red galaxies, and its errors, is shown as dashed black line. As a red dashed line, results obtained for the M$_b$/M$_{vir}$ relation in \citet{GonzalezSamaniego2014a}
 when using a semi-empirical model. The blue dashed line corresponds to the M$_b$/M$_{vir}$ relation calculated by \citet{Baldry2008}. 
As can be seen it becomes a challenge for 
$\Lambda$CDM to fit its theoretical predictions with observations.}
\label{fig:7}
\end{figure}

The total stellar mass in the virial radius of our G.3 models is M$_{*}$=6.1$-$6.2$\times$10$^{10}$~M$_{\odot}$, that is comparable
 with values
estimated for the Milky Way (4.5$-$~7.2$\times$10$^{10}$~M$_{\odot}$ in \citet{Flynn2006} and \citet{Liquia2014}). However, if we observe the relation between stellar mass and total mass (see
Fig.~\ref{fig:7}) derived from our models (blue dots) and also from observations of the MW (shadowed region) we see that most of them fall quite above the M$_*$/M$_{vir}$ predicted by
cosmological theories. The same result is observed when using data from other recent MW-sized simulations like ERIS simulation \citep{Guedes2011} or the ones of \citet{Mollitor2014}. In this work we have not studied this mismatch deeply, we leave it for the future.

\subsubsection{Stellar disk and stellar spheroid}\label{sec:6.2.3.1}

\begin{figure}
\centering
\includegraphics[scale=0.3]{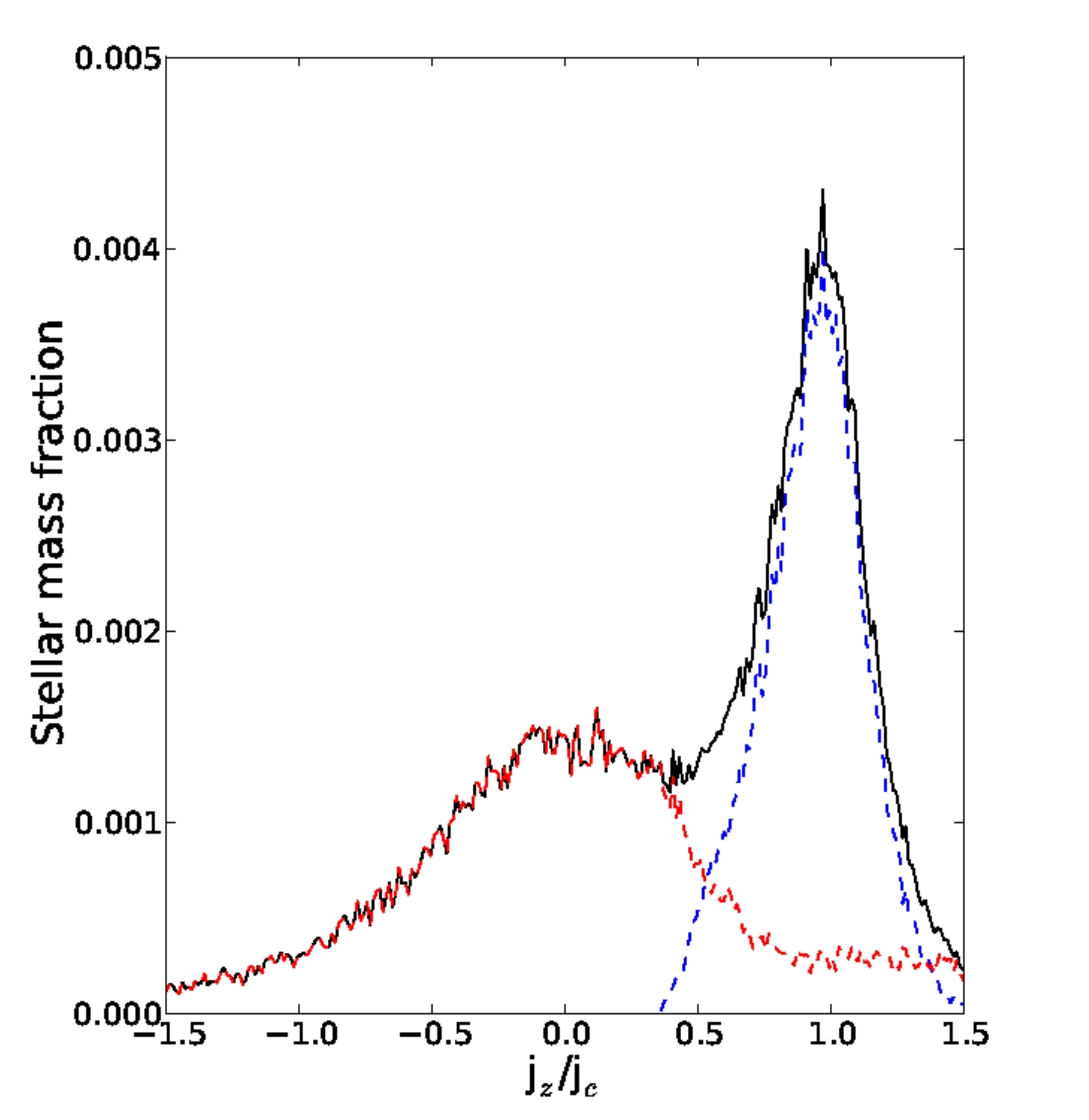}
\caption{Stellar mass fraction as function of the orbital circularity parameter j$_z$/j$_c$, describing the degree of rotational support of a given stellar particle 
(black: all particles; red: spheroid; blue: disk). The stars in a centrifugally supported thin disk manifests itself in a sharply peaked distribution 
about unity. We have plotted only stellar particles inside disk region ($\left| z \right|$~$<$~0.5~kpc and r~$<$~15~kpc) of model G.321 at 
redshift 0.}
\label{fig:8}
\end{figure}

Simulated galactic systems in our main models (G.3 series) have both stellar disks and stellar spheroidal components (halo and/or bulge). 
To study properties of stellar galactic disks we need to know which stars belong to it. To undertake the selection
process we have used an extension of the kinematic decomposition proposed by \citet{Scannapieco2009}.
 In \citet{Scannapieco2009} the authors used the ratio between the real stellar particle angular momentum perpendicular to the disk plane 
($j_z$) and the one obtained assuming stellar particles follow a circular orbit  (computed from the circular velocity curve, $j_c$). The circular velocity curve 
has been computed using both, GM/r approach and the real galactic potential; we have found that results do not depend on the v$_c$ computation technique. Using $j_z/j_c$ ratio it is possible to distinguish particles that are rotationally supported 
($j_z$=$j_c$ for disk stars) from ones that are not ($j_z$~$<$~$j_c$ for spheroid stars). Here we have decided to make additional cuts to improve this technique. We have done complementary cuts in
metallicity, in the vertical coordinate $\left|z\right|$, in the angle between the rotation axis of each particle and the vector perpendicular to the plane ($cos(\alpha)$) and 
in the distance to the galactic center (R). To avoid kinematic biases we have made cuts that do not involve kinematics first. We 
have started the selection process by making the metallicity cut; next,
 we have made a cut in the vertical coordinate, and later in radius, in $cos(\alpha)$; finally we have used $j_z$/$j_c$ condition. Some of
 these restrictions we have imposed require from a
 previous knowledge of the disk plane position. To find this position we have used an iterative geometrical
 approach \citep[pages 9 to 11 in ][]{AtanasijevicReidel}. In this approach we end up with the plane that minimizes the cumulative distance of all particles to
it. In the first iteration of this process we only use stars that are inside a  20~kpc sphere centered within the center of mass of the lightest DM mass specie particles.
 After this first
 iteration we only use stars that fall nearby the newly defined plane (i.e. $\left|z_{new}\right|$~$<$~5~kpc). We have checked that after few 
iterations we get a plane that coincides with the disk plane defined by young stars and cold gas (see bottom panels of Fig.~\ref{fig:1}, Fig.~\ref{fig:2} and Fig.~\ref{fig:6}).\\
After analyzing the stellar distribution of our G.3 models in both, Cartesian and phase spaces, we have found conditions that better distinguish disk from spheroid particles. In our
main models these conditions are:
 $j_z$/$j_c$~$>$~0.55, $\left|z\right|$~$<$~3.5~kpc, log(ZIa/ZIa$_{\odot}$)~$>$~-0.5, where ZIa$_{\odot}$~=~0.00178 and it is the
iron solar abundance according to \citet{Asplund2009},
cos$(\alpha)$ $>$ 0.7 and R $<$ 25~kpc. All particles inside r${vir}$ that do not accomplish one or more of these restrictions
 are considered spheroid particles (halo and/or
bulge particles).  To account for thick disk particles we relax the log(ZIa/ZIa$_{\odot}$) and $j_z$/$j_c$ conditions;
that is, stellar particles will also belong to the disk component if they fulfill all 
previous conditions except that they have log(ZIa/ZIa$_{\odot}$)~$<$~-0.5 and $j_z$/$j_c$
deviates by less than 0.75  from $j_z$/$j_c$ peak\footnote{For example,
if the distribution peaks at 1.0 then a particle with $j_z/j_c = 0.3$
would belong to the disk.}. The result
of this selection can be seen in Fig.\ref{fig:8} where we show the stellar mass fraction as a function of $j_z$/$j_c$ for model G.321, the result
is similar for the other models. In this figure we show how well we trace the two main stellar components of the model: one centered at $j_z$/$j_c$=1, 
i.e. rotationally supported, and another at $j_z$/$j_c$=0.

The mass of the rotationally supported component (stellar disk) is around M$_d$~=~1.82, 2.17, 2.21$\times$10$^{10}$M~$_{\odot}$ and that
of the spheroidal  one (bulge + halo) is
M$_{sp}$~=~4.29, 3.93, 3.79$\times$10$^{10}$~M$_{\odot}$ for models G.321, G.322 and G.323, respectively.
 Comparing these results with observations  we find that the
mass of the spheroidal component in our models (bulge + halo stars) 
is higher than the upper limit obtained by \citet{Flynn2006} which is M$_{b}$~$\le$~1.3$\times$10$^{10}$~M$_{\odot}$  or than the most recent observations
presented by \citet{Valenti2015}  (M$_{b}$=2.0$\pm$0.3$\times$10$^{10}$~M$_{\odot}$). The
spheroids of our galactic systems are $\sim 2$ times heavier than the one of the MW. On the other hand, disks
in our G.3 models are less massive than the one estimated for the MW by \citet{BovyRix2013} 
($\sim$4.6$\times$10$^{10}$~M$_{\odot}$). As a consequence, 
the kinematic disk-to-total ratio (D/T) decomposition, f$_{disk}$, computed as M$_{*disk}$/M$_*$ \citep{Scannapieco2010,Scannapieco2015} where M$_{*disk}$ is the mass of stellar particles
with j$_z$/j$_c$ $>$ 0.5, results in a value ($\sim$0.42) that is significantly lower than the one obtained in 
MW observations \citep[$\sim$0.75 in ][]{Scannapieco2011}. However, as pointed out by \citet{Scannapieco2010,Scannapieco2015} the
use of a kinematical
decomposition results in a lower D/T than those obtained using a photometric decomposition, as the one determined in MW 
observations. For instance, in \citet{Scannapieco2010} the authors obtained a set of MW-sized galaxies with a kinematical f$_{disk}$~$\sim$~0.2 and 
they demonstrated that f $_{disk}$ becomes 0.4$-$0.7 when using a photometric decomposition.\\

 Is the presence of this massive spheroidal component due to a large stellar concentration in the central 
region of our simulated galaxies, a common result in earlier hydrodynamical zoom-in 
simulations? As can be seen in the rotation curves shown in Fig.~\ref{fig:4} this is not the case 
(the curves do not present a central peak). We have analyzed this spheroidal component and we have detected
 that it is mainly a triaxial structure that was formed in the major merger occurred at z~=~3 (see Sec.~\ref{sec:morfologia}). This 
triaxial structure has a low density in the disk region and thus it is not a classical massive bulge.

\subsubsection{Disk properties}\label{sec:6.2.3.2}

\begin{figure}
\centering
\includegraphics[scale=0.3]{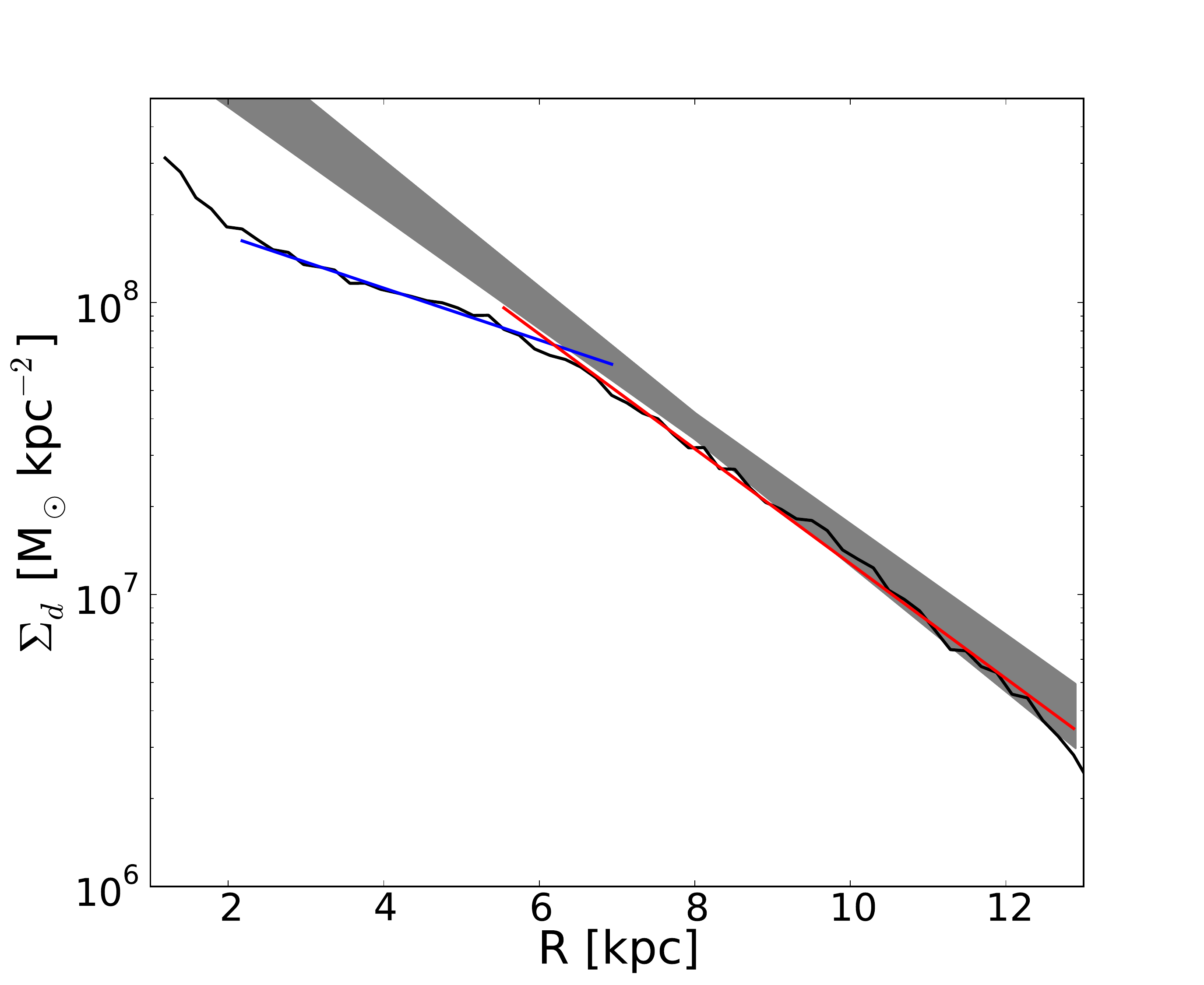}
\caption{Stellar disk surface density profile as function of cylindrical radius R (black solid line). The solid blue and red lines show the best-fit exponential profiles to the inner and outer 
disk components, respectively. The inner component (2$-$6~kpc) has an R$_d$=4.89~kpc and the outer (6$-$12~kpc) an R$_d$~=~2.21~kpc. If we fit the whole
 disk density profile from 2 to 12~kpc we obtain an R$_d$~=~2.56~kpc. All results have been obtained analyzing model G.321 at redshift 0. As a gray shadowed
region we show MW surface density profile obtained by \citet{BovyRix2013}.}
\label{fig:10}
\end{figure}

\begin{figure}
\centering
\includegraphics[scale=0.3]{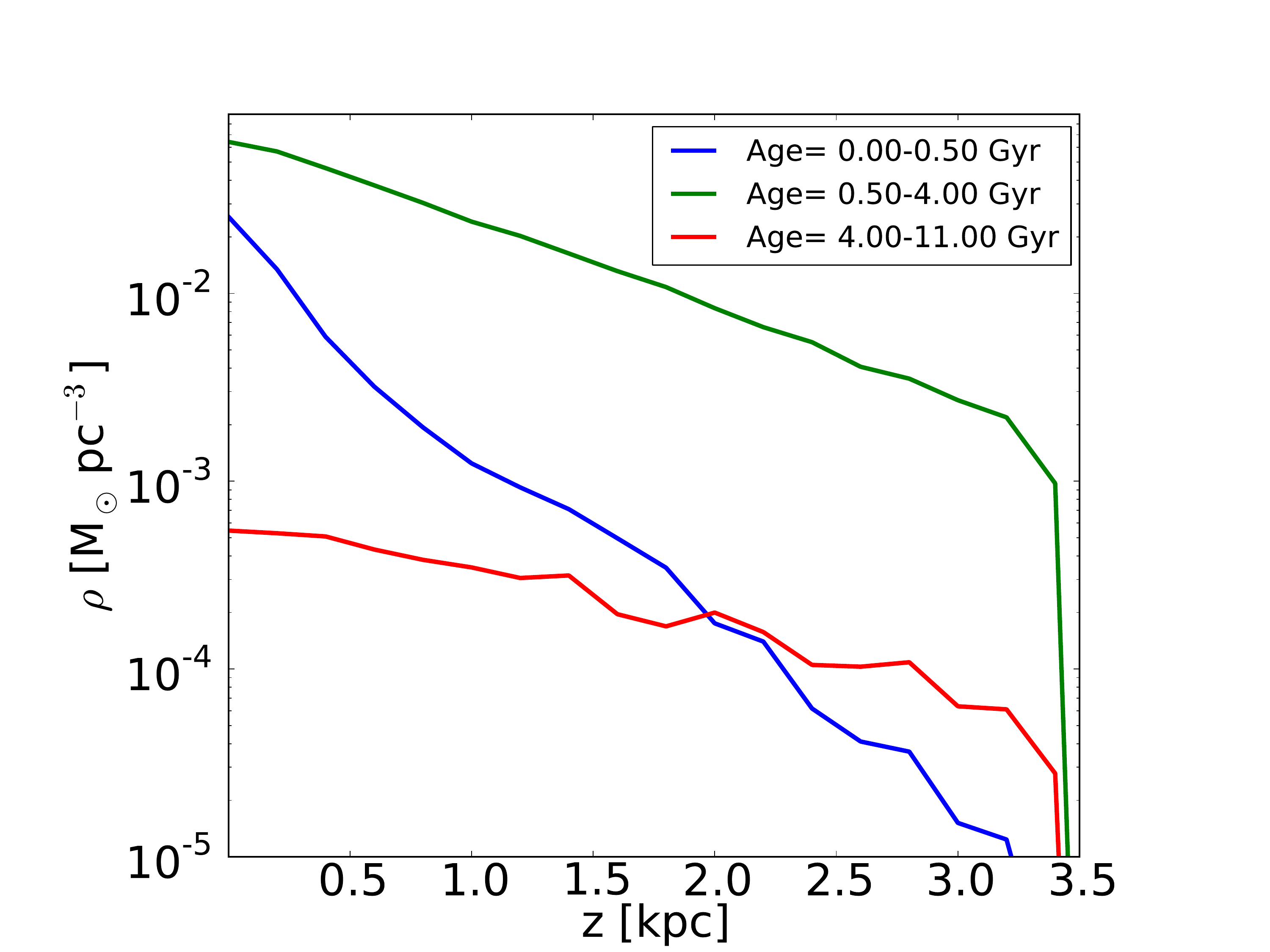}
\caption{Vertical density profile of young (blue), intermediate (green) and old (red) disk stars in the model G.321 at redshift 0. }
\label{fig:12}
\end{figure}

With the stellar disk/spheroid selection process we have found that disk component has 2.0, 6.0 and 18.7$\times$10$^5$ particles 
in each one of our main G.3 models. We have also computed the mean volume and surface 
density of the disk, as function of radius. We have found that a single or a double exponential power law can be fitted to the data depending on the age of the 
selected population. In Fig.~\ref{fig:10} we show the total disk surface density (black) as function of radius of the model G.321 at z=0.
 We have fitted two exponential power laws (red and blue) to the surface density curve. Results from these fits are that scale length in the inner regions
 (2$-$6~kpc) is R$_d$=4.89~kpc and in the outer ones (6$-$12~kpc) R$_d$=2.21~kpc.
 If we fit a single exponential to the whole radial range from 2 to 12 kpc we obtain that R$_d$=2.56~kpc, result that is in agreement with values obtained for the MW (e.g. 2.3$\pm$0.6~kpc in \citet{Hammer2007} or
2.15$\pm$0.14~kpc in \citet{BovyRix2013} using G-type dwarfs from SEGUE). It is important to note that, as can be seen in Fig.~\ref{fig:10}, a small concentration of stars is present in the central regions (R $<$ 2~kpc). This 
concentration is caused by the presence of a young bar.\\
In Fig.~\ref{fig:12} we show the volume density as function of the distance to the plane (z) of three disk stellar populations split by age. 
We have selected a young (0$-$0.5~Gyr), intermediate (0.5$-$4.0~Gyr) and old (4.0$-$11.0~Gyr) populations. The density have been
computed using vertical bins of $\Delta$z=0.2~kpc and only stellar particles that are in between R=2 and R=9~kpc.
We have found that the scale height (h$_z$) of each population,  when fitting a decreasing exponential to the vertical density distribution
in the range 0.1~kpc~$<$~$\left|z\right|$~$<$~2.0~kpc, is h$_z$=277~pc for the young, h$_z$=959~pc for the intermediate and h$_z$=1356~pc for the old populations.
These values have been obtained from G.321 model, results from models G.322 and G.323 can be found in Tab.~\ref{tab:1}. These scale 
height values are compatible with ones expected for the MW: observations show that the vertical distribution of stars in the MW 
can be fitted by two exponential with a scale heights of h$_z$=300$\pm$60~pc \citep{Juric2008} and h$_z$=600$-$1100$\pm$60~pc
 \citep{Du2006}. Our values also follow the relation proposed by \citet{Yoachim&Dalcanton2006} that argue that disk scale height in 
edge on galaxies increases in disks following z$_0\approx$2h$_z$=610(V$_c$/100)~km~s$^{-1}$.\\

\subsubsection{Star formation history (SFH)}

\begin{figure}
\centering
\includegraphics[scale=0.45]{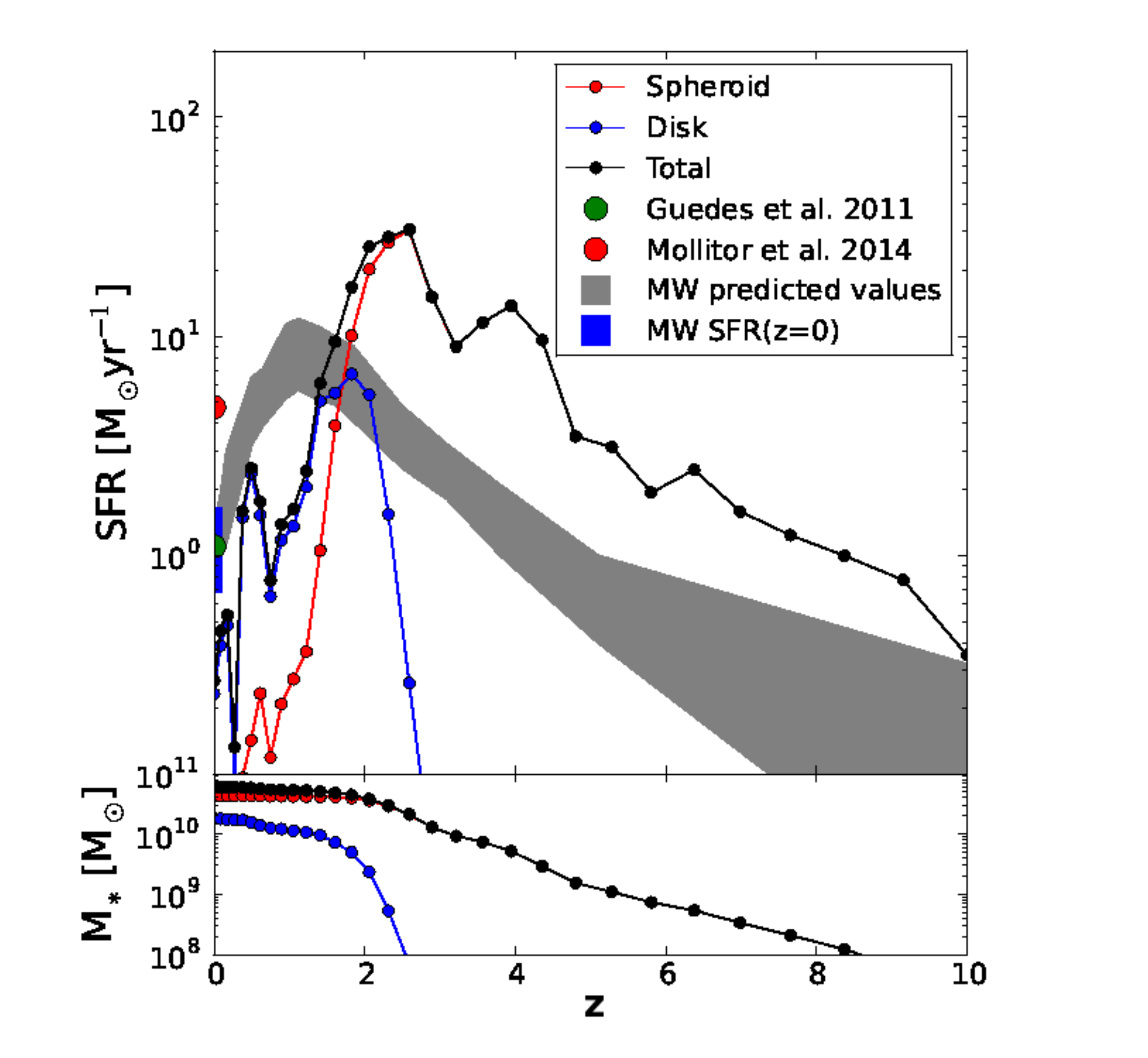}
\caption{Star formation history (SFH), top, and total mass history, bottom, of all star particles identified within virial radius at z=0 in model G.321. 
Black, blue and red filled dots show the total, disk and spheroid star 
formation rates (top panel) and stellar mass (bottom panel), respectively, as a function of redshift. Grey shadowed region indicates  the predicted
SFH for a MW-like halo with M$_{97}$=10$^{12}$~M$_{\odot}$ \citep{Behroozi2013}. Green and red big dots show total SFR at z=0 of models presented by \citet{Guedes2011, Mollitor2014} (B), respectively. The blue filled region shows the z=0 MW SFR range derived from observations by \citet{Robitaille&Whitney2010}.}
\label{fig:13}
\end{figure}

In our simulations stars are being formed with a SFR$_{z=0}$=0.27~M$_{\odot}$~yr$^{-1}$, 0.18~M$_{\odot}$~yr$^{-1}$ and 
0.41~M$_{\odot}$~yr$^{-1}$, for models G.321, G.322 and G.323, respectively. These values are smaller 
than ones inferred by 
\citet{Robitaille&Whitney2010} using Spitzer data, for the MW (SFR=0.68$-$1.45~M$_{\odot}$~yr$^{-1}$) or the one
from \citet{Liquia2014} that is 1.65$\pm$0.19~M$_{\odot}$~yr$^{-1}$. In Fig.~\ref{fig:13}, top panel, we show the SFR as function
of the redshift for the total (black), spheroidal (red) and disk (blue) components.  We have computed the SFH of each stellar component using the  
formation time of their corresponding
stellar particles (saved by the code) inside 
r$_{vir}$ at z=0. We have followed the strategy described in Sec.~\ref{sec:6.2.3.1} to distinguish between disk and spheroid 
stellar particles.
We also show, as a shadowed region, the predicted SFH for MW 
like halos derived from semi-analytical models that combine stellar mass functions 
with merger histories of halos \citep{Behroozi2013}. The peak in the SFH of our G.3 models occurs at slightly early ages than the one predicted for a MW-sized galaxy.
 In the bottom panel we plot the total stellar mass
 as function of redshift. Some important results that can be appreciated from this figure are, first, that the spheroidal component is being build at high
 redshifts while disk starts its formation at around z=2.5 ($\sim$11~Gyr from the present time), just after the last major merger that was at z=3. Second, it is also important to note that the star
 formation of spheroidal component decrease quickly after z$\sim$2.4 and become negligible at around z=0.5. The disk SFR also decreases fast in the last time instants except for some short periods of star formation
that are a consequence of the accretion of small gaseous satellites. This reduction in the star formation when reaching z=0 is a consequence of the reduction of cold gas available for star formation.
Several processes can lead to a reduction of the cold gas mass fraction. First, cold gas can be drastically consumed at higher redshifts due to a non-realistic implementation of physical parameters
that controls star formation \citep{Liang2015}. When such problem exists a large amount of old stars are present at z=0 in the inner disk region. In our G.3 models, as can be seen in Fig.~\ref{fig:8}, an old spheroid is present in the disk region, however
this old stellar population is not big enough to account for the total reduction of the observed cold gas component. Second, if stellar and SNe feedback is too 
efficient, an important amount of cold gas can be heated up and then to become unavailable for star formation. In this second scenario what we would expect is that the fraction of hot gas increase with time, a behavior that
is not observed in our G.3 models. Finally a decrease in the cold gas inflow due to inhomogeneities in the circumgalactic medium (CGM) can also be a cause of the reduction of the SFR. In this last hypothesis we would observe
only a small amount of cold gas present around our system, or falling from filaments, in the last Gyr. We argue that in our models the reduction of the 
amount of cold gas is a consequence of a combined effect of the first
and the last hypothesis as we observe that an old spheroid is present and also that there is not a large amount of cold gas falling from
filaments at z=0, however we leave the deeper study of such processes for the future.

 In ART, the metallicity $Z$ is divided according to which kind of SNe produce the metals: $ZII$ or $ZIa$, if they are 
produced by SNII or SNIa, respectively. The metallicity $Z$ is then $Z = ZIa + ZII$. Alpha elements, as oxygen, are
mostly produced by SNII while iron is mostly produced by SNIa, we thus can assume that the abundance of the former
are proportional to ZII while the iron abundance is proportional to ZIa.

\subsection{Are our G.3 models MW-like systems?}

Although it is not the aim of this paper to obtain a realistic model of the MW galaxy, 
nevertheless the models we present here 
(i.e. G.321, G.322 and G.323) can be considered at least MW-sized as they reproduce several 
of the MW observed properties. On the other hand, 
it is also evident that some parameters of these models fall far from
observations of the MW. Below, we give an overview of the parameters that match MW observations and 
of those ones that do not.\\

\subsubsection{MW-like properties}

Our study is focused on a galactic system, the G.3 series, that forms spiral arms and a 
galactic bar. Its assembly history is quiet after z=1.5 and there are no major mergers after z=3, 
something that is also expected for our Galaxy \citep{ForeroRomero2011}.
Its total mass is inside the observational
range proposed for the MW by \citet{Xue2008,BoylanKolchin2009,Kafle2012,Kafle2014} that is 1.0~$\pm$~0.4~$\times$~10$^{12}$~M$_{\odot}$. 
The stellar mass inside virial radius (6.1~$-$~6.2~$\times$~10$^{10}$~M$_{\odot}$) falls also
inside the observational range for the MW that is 4.5~$-$~7.2~$\times$~10$^{10}$~M$_{\odot}$
\citep{Flynn2006,Liquia2014}. Finally, the cold gas mass inside virial radius in model G.321 
(9.34~$\times$~10$^{9}$~M$_{\odot}$) also match the observed values
(7.3~$-$~9.5~$\times$~10$^{9}$~M$_{\odot}$ in \citet{Ferriere2001}).
Several structure parameters of our G.3 models are also within the
MW observations. 
This is the case for the dark matter halo concentration
parameter (c) that in our models takes values from 26.8 to 28.5 while observations predict it should be 21.1$^{+14.8}_{-8.3}$ \citep{Kafle2014}.
Hot gas power-law profile index ($\alpha_X$), defined in Sec.~\ref{sec:6.2.4}, 
and that in our G.3 models is in between -0.62 and -0.83, is also close to value proposed by \citet{Anderson&Bregman2010} which
is -0.9. Disk scale lengths and heights of our galactic systems (G.3 series) 
also coincide or are close to observed values. The former in our models
is 2.56$-$3.2~kpc and observed values are 2.3$\pm$0.6~kpc in \citet{Hammer2007} and 2.15$\pm$0.14 in \citet{BovyRix2013}. Observed disk
scale height of young and old stars is 300~$\pm$~60~pc in \citet{Juric2008} and 600$-$1100~pc in \citet{Du2006}, respectively, and in
our models we have obtained for young and old populations, 277$-$393~pc and 960$-$1356~pc, depending on the model.
Finally, also several kinematical parameters agree with the MW
observations; the rotation curve of our simulated galaxies, for instance, roughly match
observations (see Fig.~\ref{fig:4}). Some examples of such agreement are the ratio V$_{2.2}$/V$_{200}$ that is 1.9 in our models and 
1.67$^{+0.31}_{-0.24}$ in \citet{Xue2008}, and the v$_c$(R$_{\odot}$) that is 233.3$-$239.8~km~s$^{-1}$ in this work and 
221$\pm$18 in \citet{Koposov2010} or 236$\pm$11 in \citet{Bovy2009}.

\subsubsection{Parameters that do not match observations}

Among the parameters that do not agree with the observed values
of the MW is the SFR and the bulge to disk ratio. The cold gas fraction in the G.323 and G.323
models also do not match observations, they account for only  1.52 and 
1.86~$\times$~10$^9$~M$_{\odot}$ while observed values are 7.3~$-$~9.5~$\times$~10$^9$~M$_{\odot}$ 
\citet{Ferriere2001}. On the other hand, the SFR in 
our models ranges from 0.18 to 0.41~M$_{\odot}$~yr$^{-1}$ values below the observational range 0.68~$-$~1.45~M$_{\odot}$~yr$^{-1}$
from \citet{Robitaille&Whitney2010}.
Finally, we have presented in Sec.~\ref{sec:6.2.3.1} the stellar spheroid-disk decomposition and  
shown that a massive spheroid 
exists in our models. Such massive structure has not been observed in the MW. The total mass of the
spheroid (bulge+halo) and the stellar disk in our G.3 models are 
3.79$-$4.29~$\times$10$^{10}$~M$_{\odot}$ and 1.82$-$2.21~$\times$10$^{10}$~M$_{\odot}$,
respectively. Moreover,
observations show that the bulge and the disk mass of our Galaxy are $\le$2.3$\times$10$^{10}$~M$_{\odot}$ \citep{Flynn2006,Valenti2015} and
$\sim$4.6$\times$10$^{10}$~M$_{\odot}$ \citep{BovyRix2013}, respectively; that is, the spheroid of
our models is at least two times more massive than the bulge of the MW.

\section{Missing baryons problem and the X-ray luminous hot gas}\label{sec:results1}

\subsection{The missing baryons problem} 

It was long been known that the cosmological baryon fraction inferred from Big Bang nucleosynthesis is much higher
 than the one obtained by counting baryons at redshift z=0 \citep[e.g.][]{Fukugita1998}. However, it was recently, with WMAP and Planck  high precision data of the cosmic microwave background \citep{Dunkley2009,Planck2013}, that the cosmic baryonic fraction was well constrained and thus the lack of baryonic mass in galaxies became evident.
This data revealed that the cosmic ratio between baryonic and total matter ($\Omega_b/\Omega_M$) is 3$-$10 times larger than the one observed in galaxies.
For instance, \citet{Hoekstra2005} found a baryonic fraction for isolated spiral galaxies of 0.056 and for elliptical of 0.023 while 
\citet{Dunkley2009} presented a cosmic baryonic fraction from WMAP 5-year data of about 0.171$\pm$0.009.\\
Studies like \citet{Cappi2013,Georgakakis2013} tried to solve this missing baryon problem proposing that galactic winds, SNe 
feedback or strong AGN winds ejected baryons to the circumgalactic medium (CGM). Others proposed that most of the gas never collapsed 
into the dark matter halos as it was previously heated by SNe of Population III \citep[e.g. ][]{Mo&Mao2004}, this is known as the preheating scenario. In this last scenario missing 
baryons are still in the extragalactic warm-hot intergalactic medium (WHIM), following filaments.
Thus, it has become clear that studying the CGM is 
necessary not only for finding the missing baryons, but also to understand its properties in the context of galaxy formation and 
evolution models. To further constrain galaxy evolution and find out which one of the proposed theories solves
 the missing baryon problem it is needed to measure the amount of warm-hot gas phase present in the CGM and also its 
metallicity. Results of such measures give information about the heating mechanism: hot gas with low metallicities is indicative of
 a Pop III SNe preheating while a more metallic
 gas suggest that it has been enriched in the disk via SNe and stellar winds.\\
Supporting the first hypothesis, the idea of very strong feedback driving galactic-scale winds,
 causing metal-enriched hot gas to be expelled out of the star-forming disk possibly to distances comparable to or beyond the virial radius was
generally accepted.
 This process also lies in the heart of obtaining realistic simulations of disk galaxies \citep{Guedes2011,Marinacci2014}. By this mechanism gas is temporally
 placed far from star forming regions delaying the star formation and preventing the formation of an old bulge in the center of the system.\\
 From the analysis of OVII and OVIII X-Ray absorption lines observed towards the line of sight of extragalactic sources, several authors performed estimations of the hot gas mass in the MW halo, suggesting that an important part of the missing baryons is located in this hot gas phase \citep{Gupta2012,Gupta2014}. They concluded that the warm-hot phase of the CGM is extended over a large region around the MW, with a radius a mass around 100~kpc and 10$^{10}$~M$_{\odot}$. However, these estimations are limited to a few directions towards extragalactic sources (bright QSOs) and as a consequence issues like the homogeneity or isotropy of the hot gas distribution possibly affecting the mass estimation are hard to be addressed by them. Other studies like  \citet{Feldmann2013} also suggest that the presence of this hot gas phase can explain part of the isotropic gamma-ray background observed in Fermi Gamma-ray Space Telescope.\\

\subsection{Hot gas component in G.321} 
 \subsubsection{Spatial distribution}\label{subsubsec:4.2.2}
The amount of hot gas in our G.3 models is in between M$_{hot}$~=~0.98$-$1.32$\times$10$^{10}$ M$_{\odot}$.
 This hot gas that has a temperature above 3.0$\times$10$^5$~K (has opacity in the X-rays) is embedded in the dark matter halo but mostly outside the stellar disk. 
 Aiming to perform a fair comparison with observations we estimated the hot gas Hydrogen equivalent column density and emission measure (EM)\footnote[1]{
In this work we are not showing the dispersion measure values (DM) as they only differ by a factor of two from the
N$_H$ ones. N$_H$ values are presented in Fig.~\ref{fig:16}. Although not presented here, DM values in our models are of the same
 order as the ones in \citet{Guedes2011}.} as:
\begin{equation}\label{eq:6.1}
 N_H = \int_{los} n_H dl
\end{equation}
\begin{equation}\label{eq:6.3}
 EM = \int_{los} n_e^2 dl
\end{equation}

where n$_H$ is the hydrogen particles density, dl is an element of the path in the line of sight and $n_e$ is the electron density.\\

In Fig.\ref{fig:16} we show a full sky view of hot gas column density distribution in our model G.321. This figure has been obtained computing the hot gas column density from a position 
 that is at 8~kpc from the galactic center, inside the simulated galactic disk, resembling the Sun position,
 and assuming arbitrary azimuthal angles.
We have obtained column density values that fall near the observational ranges, if a solar metallicity is assumed  \citep[see Tab.~\ref{tab:4}
 and][]{Gupta2012}. However, when assuming lower metallicities, observations are consistent with values that are above the ones measured in  our simulations. This change can give us information about the absorption of the local WIMP.
We also conclude that the distribution of hot gas inferred from all estimations is far from homogeneous. In order to quantify the hot gas anisotropy we followed two strategies. First we computed the amplitude of the first spherical harmonics (Y$_l^m$ from l,m=0 to 5 ) and later we also computed the filling factor as defined by \citet{Berkhuijsen1998}. Our results show that the dominant spherical harmonic is Y$_1^0$, that is an indicator of the dipolar component, with a small contribution of several high order components.  The asymmetry of the hot gas distribution with respect to the disk plane suggest some degree of interaction with the
extragalactic medium. The computation of the mean filling factor from line of sights distributed all along the sky gives as a value of f$_v\sim$0.33$\pm$0.15. This value for the filling factor also suggests  that hot gas is concentrated in a few regions of the sky.
It is also
important to mention that  in our G.3 models (see Fig.~\ref{fig:6} and \ref{fig:16}) we do not detect a hot gas thick disk component.
 This is important
as it is under discussion what is the contribution of the hot gas thick disk to the observed X-ray emission/absorption \citep{Savage2003}.
Additionally to the spatial distribution we study the metallicity and velocity components of hot gas cells. Metallicity and velocity fields provide us information about how the hot gas component interacts with the  extragalactic medium.
 In Fig.~\ref{fig:17} we show the metallicity distribution of the hot gas as seen from the Galactic center. It is evident that gas metallicity distribution does not depend on the
 azimuthal angle ($\theta$) (Fig.~\ref{fig:17}, top panel) while it is clear the dependency on the 
 vertical latitud ($\phi$) (Fig.~\ref{fig:17}, bottom panel). 
It is also noticeable a low metallicity component  located at the northern hemisphere that is absent at the southern, where metallicity is in average much higher.
It is important to note that in Fig.~\ref{fig:17}, bottom panel, we can distinguish a clear break in the hot gas distribution at low
latitudes. This hole coincides with the position of the stellar/cold gas disk.\\
 In order to shed some light on the bimodal origin in the metals distribution, we have analyzed a set of three color maps showing the projected metallicity
 and velocities along each one of the principal planes (see Fig.~\ref{fig:19} and Fig.~\ref{fig:20}). Fig.~\ref{fig:19} (X-Y plane, i.e. disk plane) shows that in general hot gas in the disk plane rotates following
the disk rotation direction. Fig.~\ref{fig:20} (X-Z plane, top and Y-Z, bottom) shows a more complex scenario with several vertical motions and metallicity gradients, it becomes clear that both low and 
high metallicity vertical flows are present in the hot gas halo.
 It is also clear from this figure that a bimodality in metallicity exists in the vertical direction and also that low-metallic hot gas flow has a mean motion that brings
it from outside to inside the system, i.e. it is falling from the IGM. High-metallic hot gas flows observed in the 
southern hemisphere have a slower and more irregular motion than ones in the northern. While it is easy  to understand and interpret why 
low-metallic hot gas is falling inside the system, and why in some cases hot metallic gas departs from the stellar disk region (SN feedback, stellar winds...), it is not so obvious why 
some clouds of such hot metallic gas, with sizes of tens of kpc, behaves differently. We suspect that this last high-metallic
component could be associated with a small gaseous satellite that is passing through the system at z=0 (see Fig.\ref{fig:21}). To confirm all these hypothesis we have 
made the same velocity-metallicity maps as the ones in Fig.~\ref{fig:19} and Fig.~\ref{fig:20} for several snapshots back to z~=~0.5. From the analysis of these snapshots we have confirmed that in some instants hot gas with enhanced metallicity departs from the disk
 due to feedback of stellar component. We have also observed that low metallicity gas appears coming from the CGM. Finally, at around z=0, we have seen that a gaseous satellite approaches to the main
 galaxy perturbing hot gas flows. The perturbation of the hot gas distribution by accretion of small satellites is an interesting process and it will deserve a more detailed study.
The case of LMC has been used as a constrain to the hot gas distribution \citep{Salem2015}. Finally we have also compared the disk plane gas mean metallicity with that one of the hot gas in the halo and we have found that both components have similar
 mean metallicities that is around -0.64~dex 
(except for the high and low metallicity flows discussed above).\\
A conclusion from the metallicity and kinematic analysis is the presence of hot gas flows from the 
IGM, either from the low metallicity environment of from sinking metal enriched satellites. We also conclude that the disk has a low hot gas concentration as a consequence of SNe explosions and 
stellar winds pushing hot gas up to the halo. Such results are coherent with the \citet{Dekel2006}
proposal of cold gas flows feeding halos with M$_{vir}<$10$^{12}$~M$_{\odot}$.\\

\subsubsection{Halo virial mass and the total hot gas in galaxies}

\begin{figure}
\centering
\includegraphics[scale=0.3]{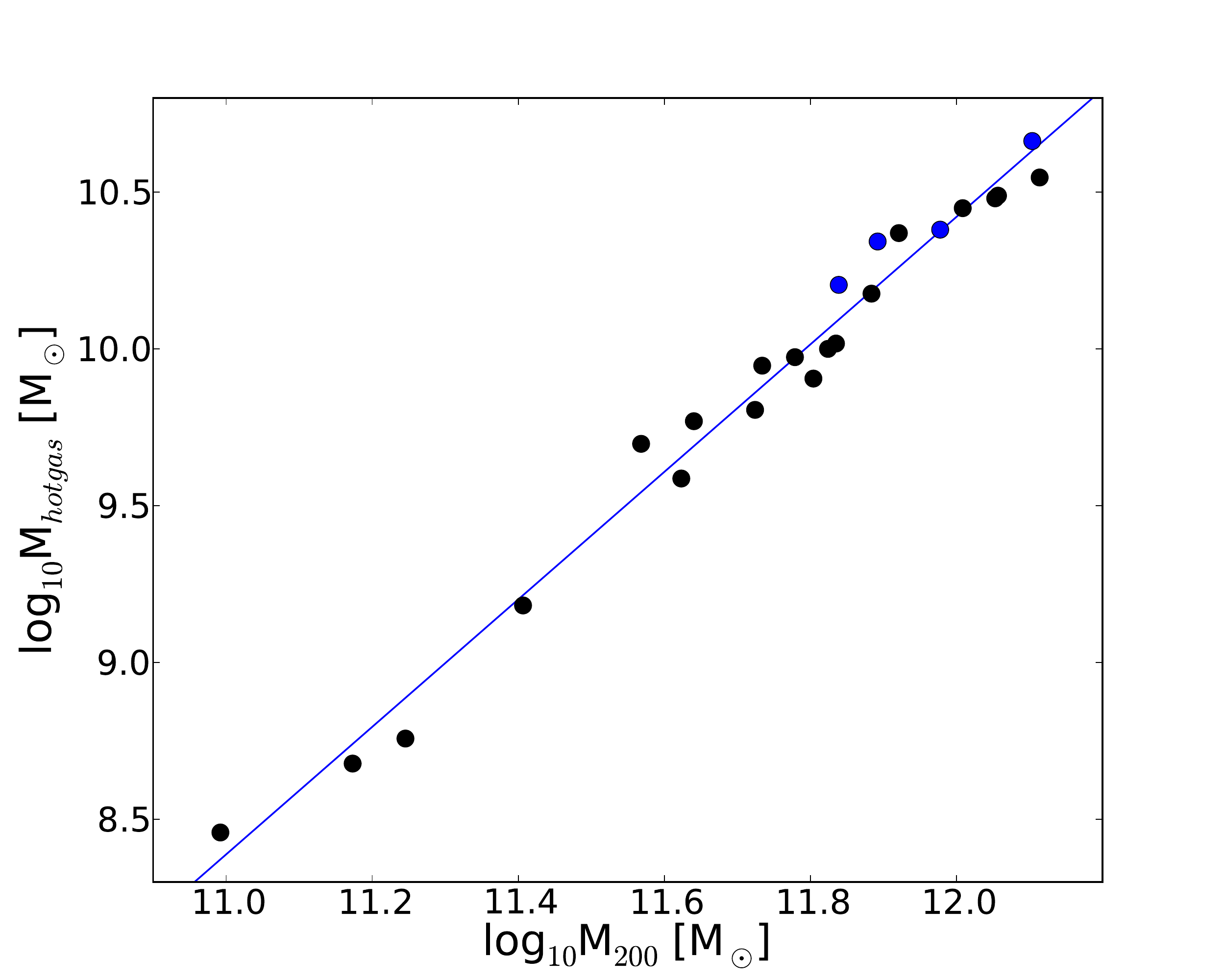}
\caption{Total hot gas mass inside r$_{vir}$~=~r$_{200}$ as function of total virial mass. Black dots: Simulations from GARROTXA run, their properties are presented in 
Tab.~\ref{tab:2}. Blue dots: Simulations outside GARROTXA run, parameters from these simulations are not presented in Tab.~\ref{tab:2}.}
\label{fig:22}
\end{figure}

In Tab.~\ref{tab:2} and Fig.~\ref{fig:22} we present results suggesting  a possible correlation between halo virial mass
 and the total hot gas in galaxies.
 After analyzing MW-sized models but also few more massive and some lighter models, we have found that M$_{vir}$ correlates with total hot gas mass. Blue dots are simulated galaxies that come from a work in preparation. These simulations will be discussed in an extended study in preparation, but we show them to illustrate the correlation persistence regardless of the subgrid
 physics. Their corresponding host halos were drawn from a 50~Mpc~h$^{-1}$ boxside
and chosen to be relatively free of massive companions  inside a sphere of 1~Mpc~h$^{-1}$. A similar result was presented in \citet{Crain2010}. In their work they found instead a correlation between X-ray luminosity and virial mass. More specifically they reported two linear correlations, one for M$_{vir}$~$<$~5$\times$10$^{12}$ and another for M$_{vir}$~$>$~5$\times$10$^{12}$. 
Our study is mainly focused on MW-size galaxy systems i.e. their low halo mass fit, consistently we have found a simple linear correlation (see Fig.~\ref{fig:22}).
Although absolute values obtained in our work can not be directly compared with the ones in\citet{Crain2010} 
as we are using M$_{hotgas}$ and they are using L$_X$, if finally confirmed this linear correlation will provide a new constrain  to virial mass in galaxies, including the Milky Way.

\begin{table*}
    \tabcolsep 4.pt
\begin{tabular}{lccccccccccc}
\hline
\hline
  Model     & Halo \# & z    & M$_{200}$               &  r$_{200}$ &  M$_{star}$           & M$_{gas}$             &M$_{hotgas}$          & M$_{coldgas}$          &  $\epsilon_{SF}$     & DM$_{sp}$ & n$_{ref}$ \\
            &         &      &  [10$^{11}$ M$_{\odot}$] &  [kpc]    & [10$^{11}$ M$_{\odot}$]&[10$^{10}$ M$_{\odot}$]&[10$^{9}$ M$_{\odot}$]& [10$^{9}$ M$_{\odot}$] &                      &           &       \\
\hline 
G.240 & 1  & 1.0  &  8.34   & 183.05     &1.31        & 3.64      &  23.4                &      12.6    & 0.10   & 3   & 11   \\          
G.240 & 1  & 0.25 &  11.3   & 203.09     & 1.63       & 4.78      &  30.2                &      17.2  & 0.10   & 3   & 11\\          
G.240 & 1  & 0.0  &  13.0   & 211.71     &1.75        &  6.86     & 35.2                 &       18.8  & 0.10   & 3   & 11\\ 
G.241 & 1  & 0.67 &  8.94   &  186.89    & 1.40       &  3.60     &  21.7                &       11.3   & 0.50   & 3   & 11   \\   
G.242 & 1  & 1.0  &  7.65   & 179.28     & 1.04       & 2.76      & 15.0                 &       10.6 & 0.70 & 4 & 10 \\         
G.242 & 1  & 0.25 &  10.2   & 198.91     &1.12        & 3.63      & 28.1                 &       6.33 & 0.70 & 4 & 10 \\                
G.242 & 1  & 0.0  &  11.4   & 207.35     & 1.15       & 5.23      & 30.8                 &       1.63 & 0.70 & 4 & 10 \\              
G.250 & 3  & 1.0  &  2.55   & 128.57     & 0.14       & 0.95      & 1.52                 &       6.24 & 0.70 & 4 & 11 \\         
G.250 & 3  & 0.25 &  4.37   &  151.82    & 0.18       & 2.63      & 5.88                 &       16.0 & 0.70 & 4 & 11 \\         
G.250 & 3  & 0.0  &  5.30   & 164.98     & 0.20       & 2.23      & 6.39                 &       12.4 & 0.70 & 4 & 11 \\         
G.260 & 4  & 1.0  &  0.982  & 92.20      & 0.03       &  0.75     & 0.287                &       6.17 & 0.5 & 3 & 11\\         
G.260 & 4  & 0.25 &  1.49   & 106.64     & 0.04       &  0.73     & 0.476                &      5.79 & 0.5 & 3 & 11\\          
G.260 & 4  & 0.0  &  1.76   & 113.50     & 0.04       &  0.80     & 0.572                &      5.49 & 0.5 & 3 & 11\\          
G.320 & 5  & 1.0  &  4.20   & 145.64     & 0.60       & 8.83      & 3.86                 &        4.10 & 0.60 & 4 & 11 \\        
G.320 & 5  & 0.25 &  6.01   &  168.45    & 0.63       &  1.18     &  9.41                &       1.26 & 0.60 & 4 & 11 \\         
G.320 & 5  & 0.0  &  6.67   & 175.59     & 0.64       &   1.46    &  10.0                &       0.405 & 0.60 & 4 & 11 \\        
G.321 & 5  & 1.5  &  3.70   & 139.71     & 0.54       &  1.33     &  4.98                &       6.17 & 0.65 & 5 & 11 \\         
G.321 & 5  & 0.67 &  5.42   &161.59      & 0.58       & 2.66      &  8.84                &       8.99 & 0.65 & 5 & 11 \\    
G.321 & 5  & 0.25 &  6.37   & 171.98     & 0.61       & 1.78      &  8.04                &       7.91 & 0.65 & 5 & 11 \\        
G.321 & 5  & 0.0  &  6.84   & 175.59     & 0.61       & 2.17      &  10.4                &       9.34 & 0.65 & 5 & 11 \\         	
\hline
\end{tabular}
\centering
\caption{M$_{200}$ vs. hot gas mass inside R$_{200}$ in our set of N-body plus hydrodynamics simulations. We also show the most relevant changes in the initial parameters,
 from the ones used in the model G.321 (see Tab.~\ref{tab:1} for the definition of the parameters).}
\label{tab:2}
\end{table*}

\subsubsection{Total hot gas mass: are results from observations biased?}

\begin{figure}
\centering
\includegraphics[scale=0.25]{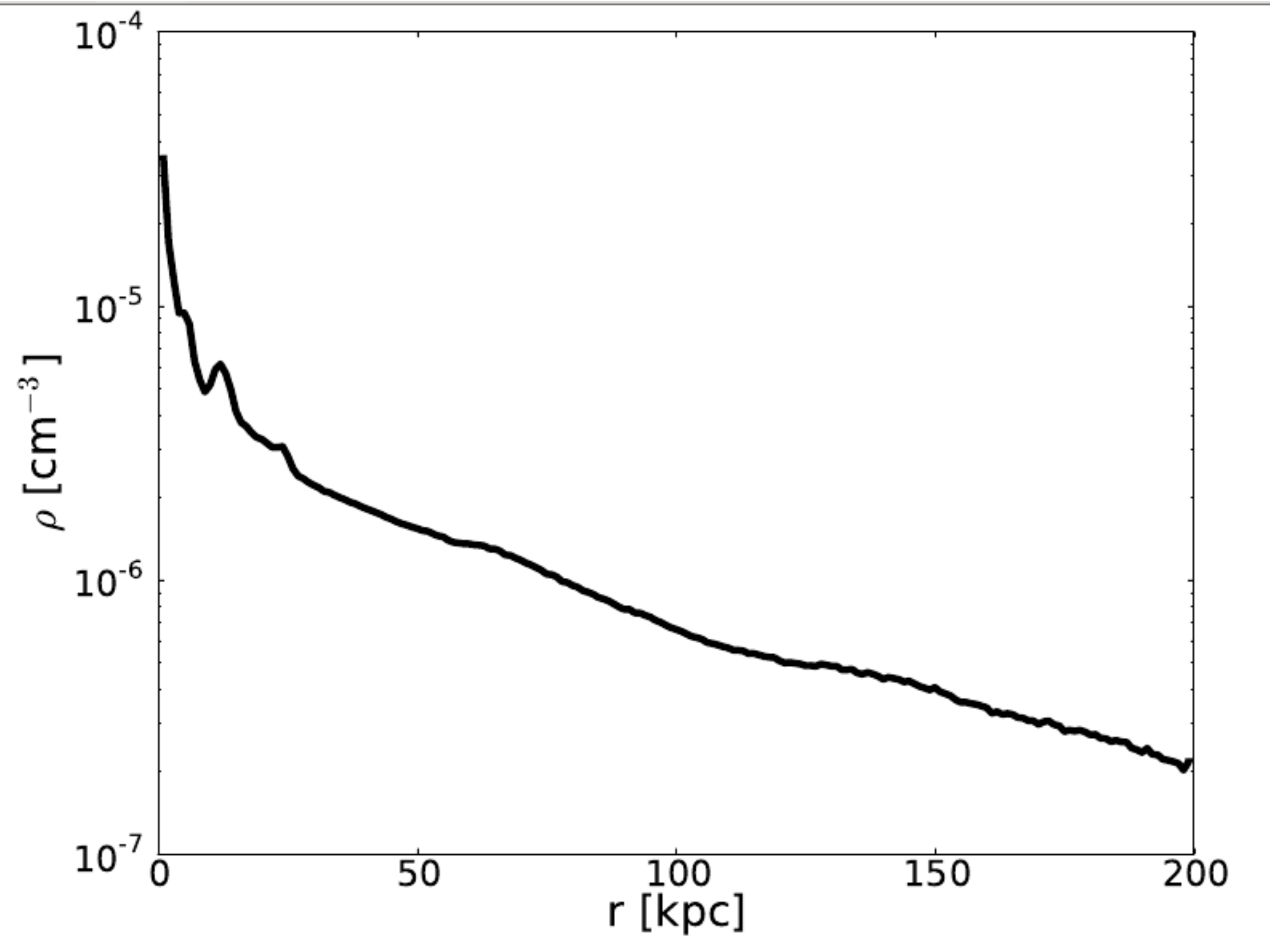}
\caption{Spherically averaged hot gas density (T~$>$~3$\times$10$^5$~K) as function of radius for the model G.321 at z~=~0.}
\label{fig:dens}
\end{figure}

\begin{figure}
\centering
\includegraphics[scale=0.17]{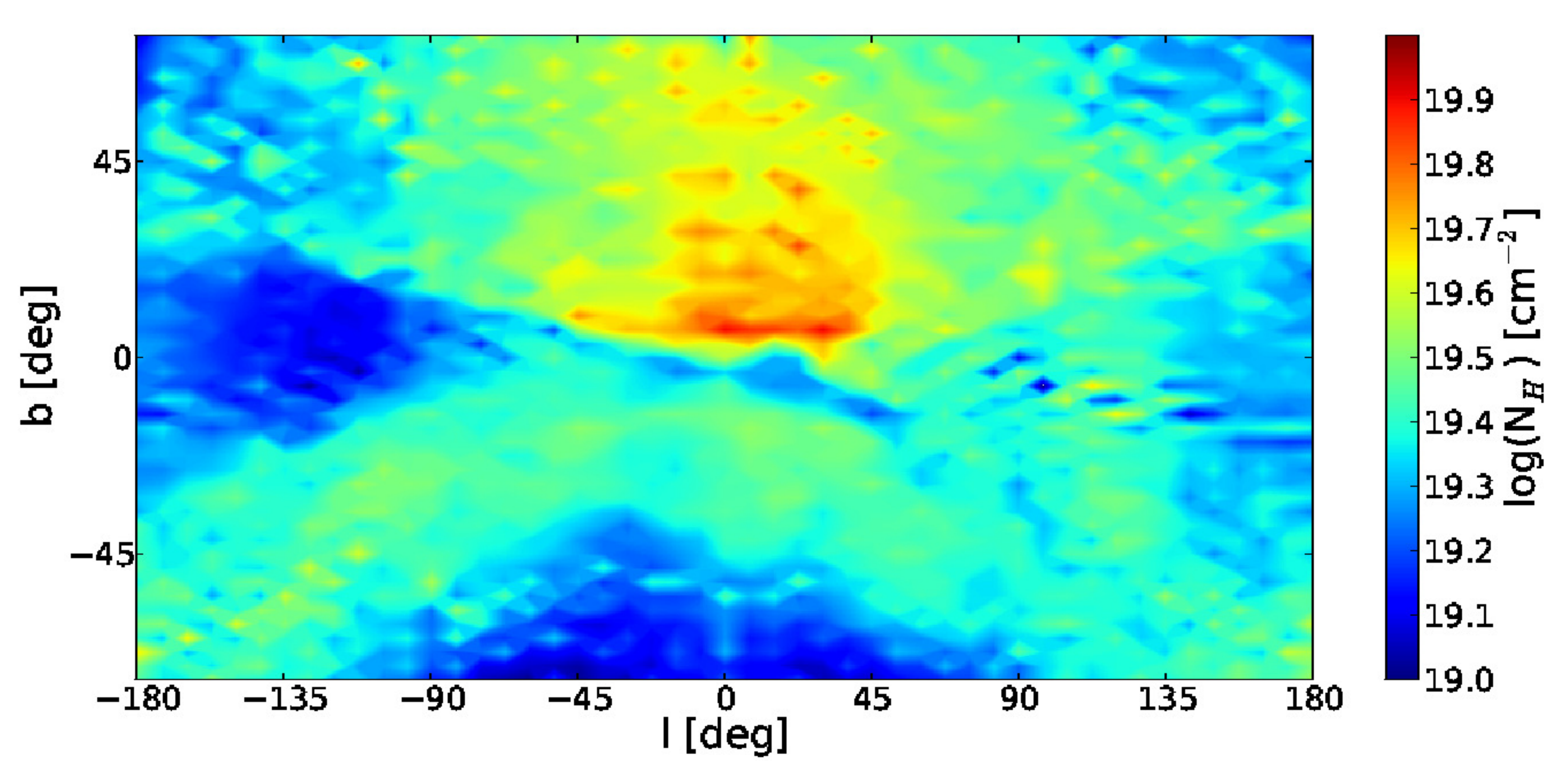}
\caption{Hot gas (T~$>$~3$\times$10$^5$~K) column density (top) in a full sky view of simulation G.321 in galactic coordinates, at z~=~0. All values have been computed as observed from R~=~8~kpc and at an 
arbitrary azimuthal angle.}
\label{fig:16}
\end{figure}

\begin{figure}
\centering
\includegraphics[scale=0.4]{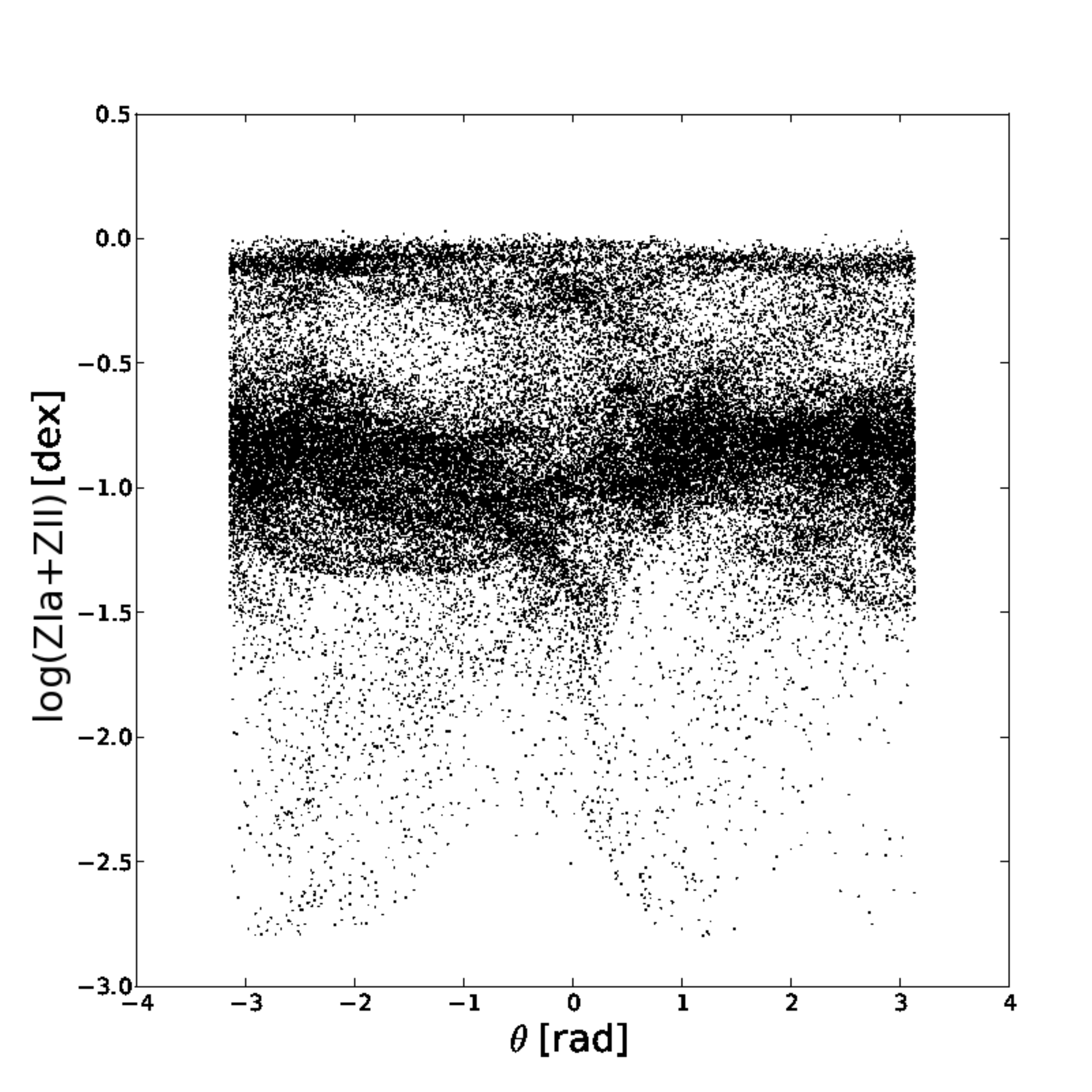}
\includegraphics[scale=0.4]{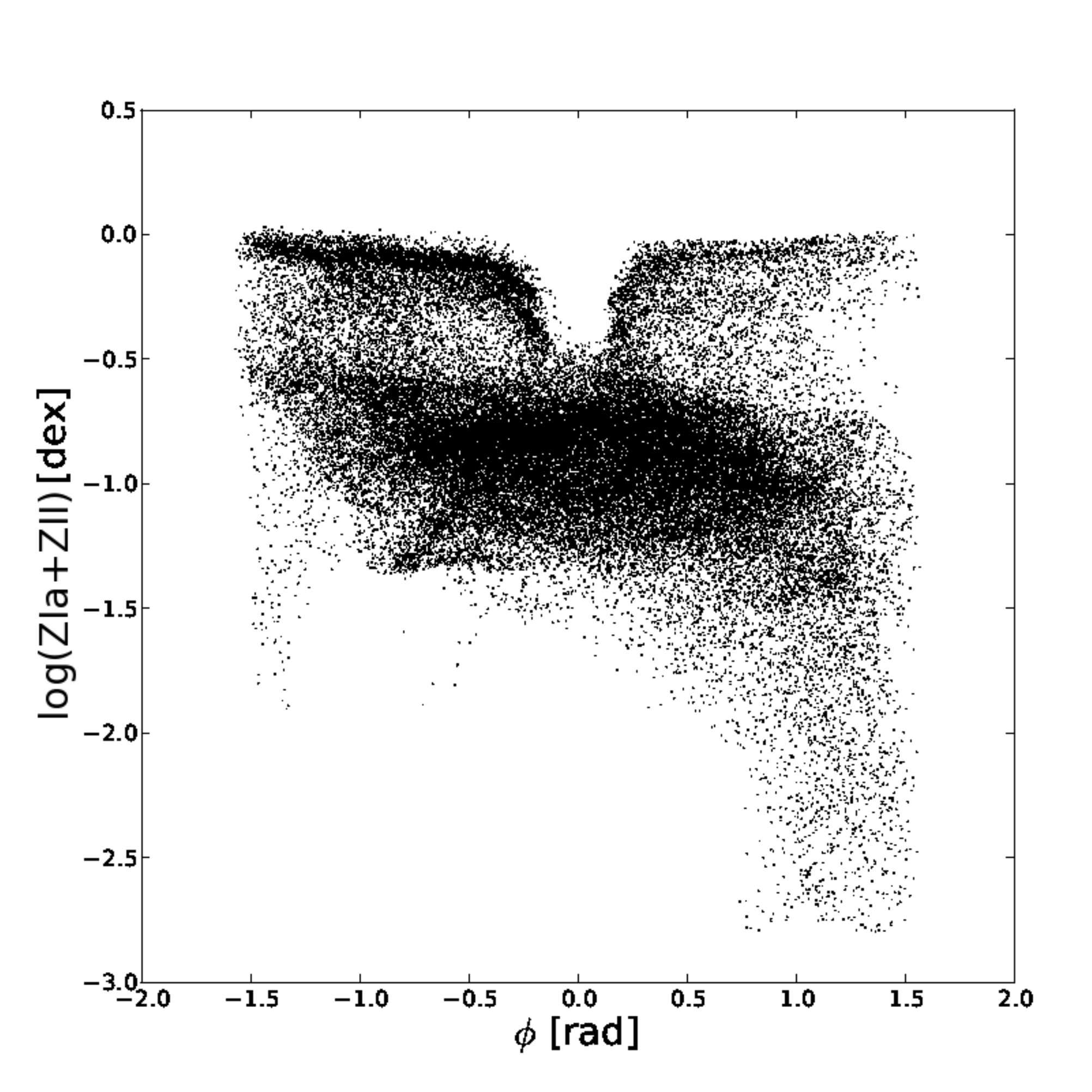}
\caption{Hot gas (T~$>$~3$\times$10$^5$~K) view in spherical galactic coordinates of the G.321 model, at z~=~0. Top: metallicity  as function of the azimuthal angle. Bottom:
metallicity as function of the vertical angle. ZIa and ZII are metals from SNe Ia and II, 
respectively. All values have been computed as observed from the galactic center.}
\label{fig:17}
\end{figure}

\begin{figure}
\centering
\includegraphics[scale=0.3]{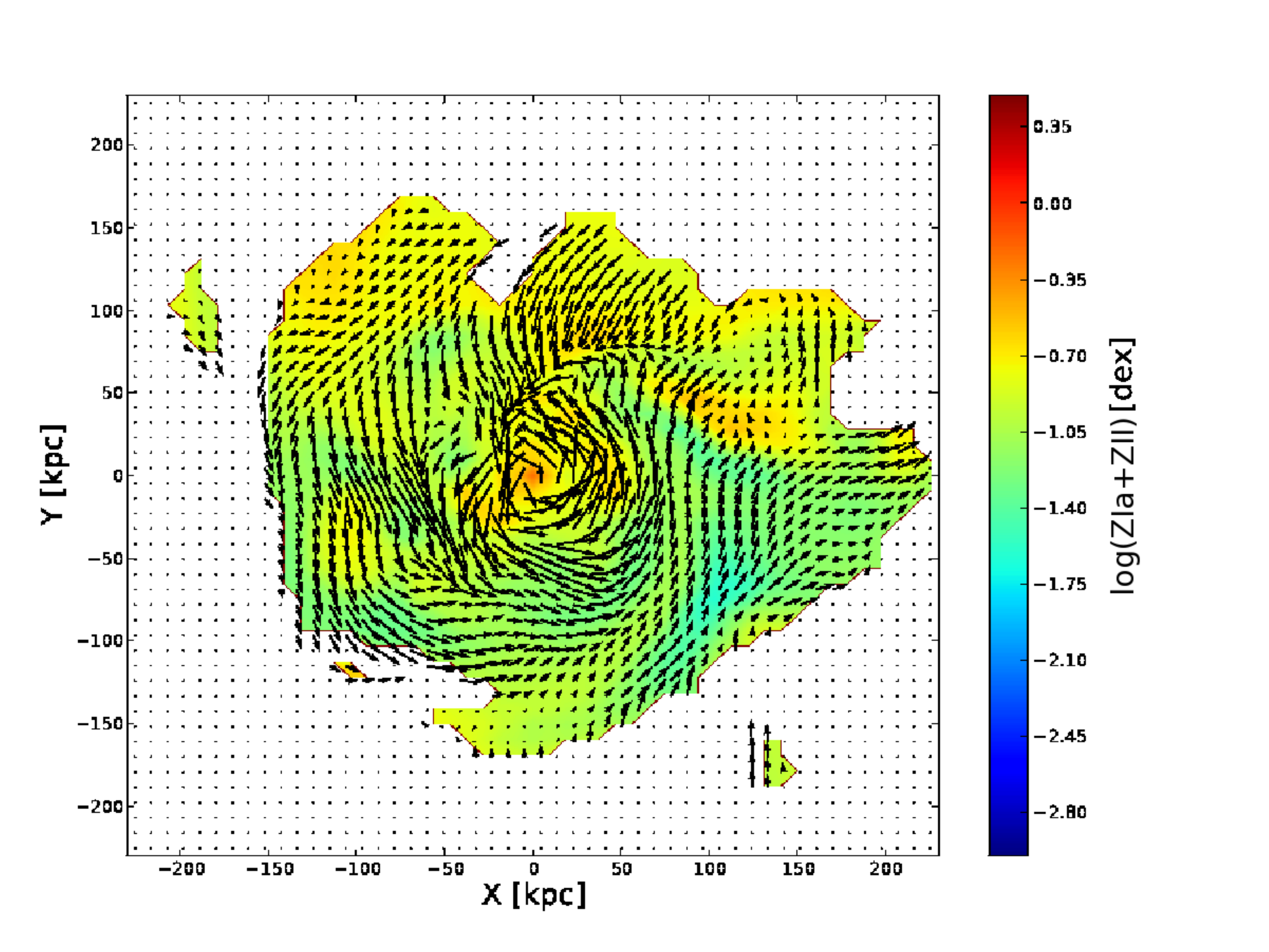}
\caption{Hot gas (T~$>$~3$\times$10$^5$~K) projected metallicity (color map) and galactocentric velocity direction (arrows) in the principal plane X-Y of the G.321 model, at z=0.
ZIa and ZII are metals from SNe Ia and II, respectively}
\label{fig:19}
\end{figure}

\begin{figure}
\centering
\includegraphics[scale=0.3]{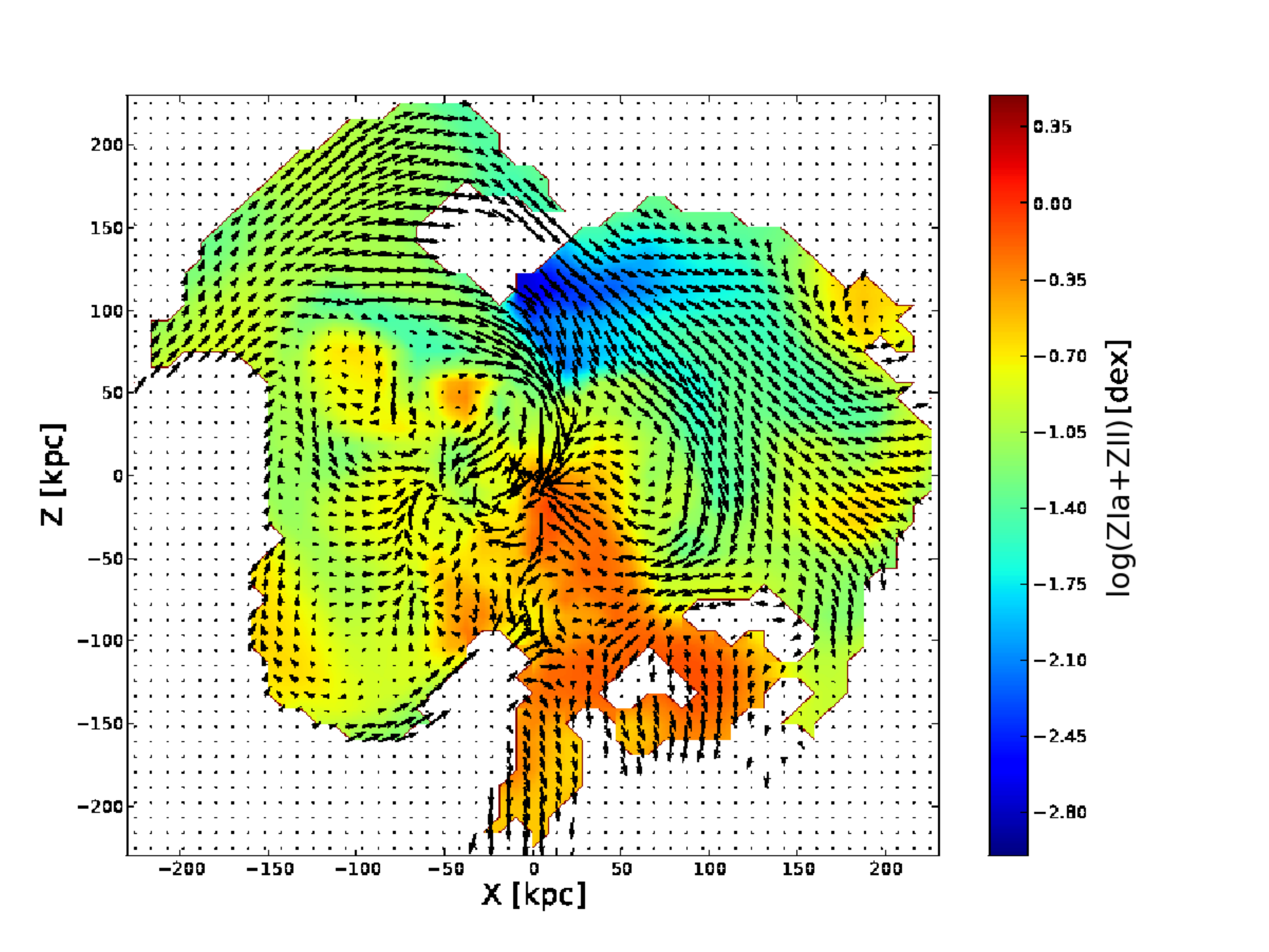}
\includegraphics[scale=0.3]{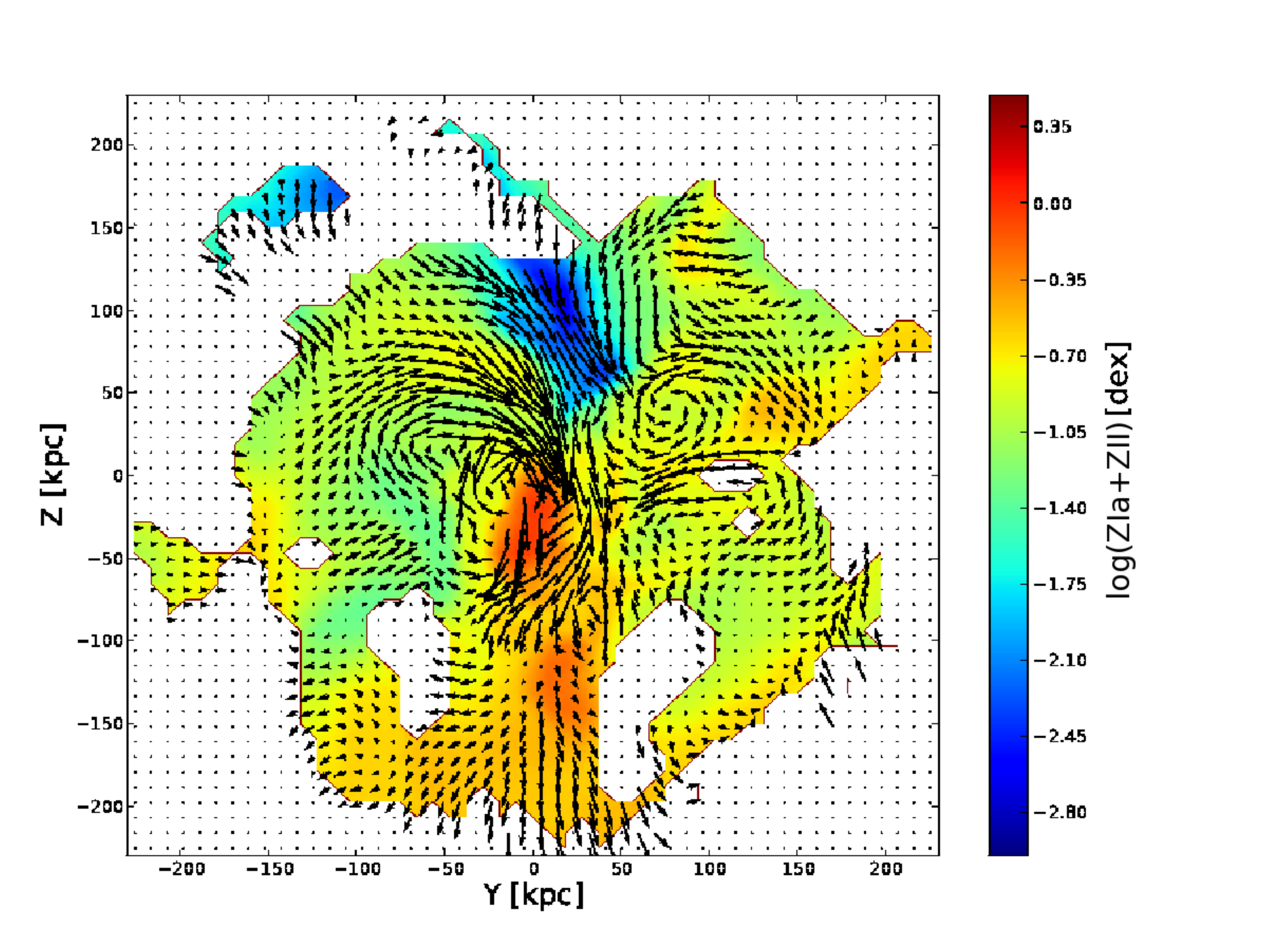}
\caption{Hot gas (T~$>$~3$\times$10$^5$~K) projected metallicity (color map) and galactocentric velocity direction (arrows) in two principal planes, X-Z (top) and
Y-Z (bottom) of the G.321 model, at z~=~0. ZIa and ZII are metals from SNe Ia and II, respectively}
\label{fig:20}
\end{figure}

\begin{figure*}
\hspace{-1cm}
\centering
\includegraphics[scale=0.6]{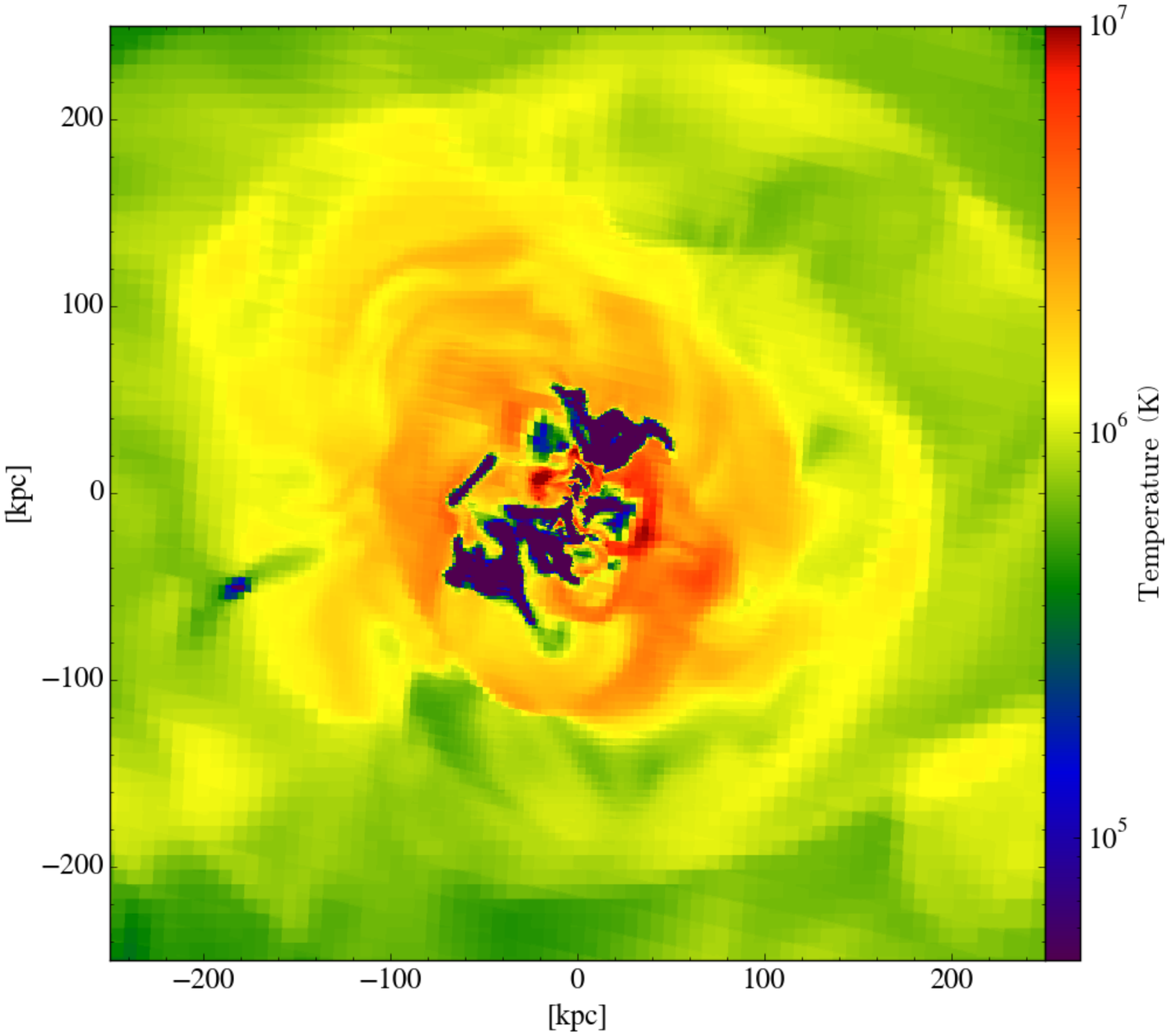}
\caption{Warm-hot gas slice plot at an arbitrary plane of the simulation G.323, at z~=~0.}
\label{fig:21}
\end{figure*}

As previously mentioned it is challenging to infer the
total hot gas mass in the MW by using the available
observational data. Matter in the CGM is hot (T~$=$~10$^5-$10$^7$~K) and tenuous (n~$\simeq$~10$^{-5}-$10$^{-4}$~cm$^{-3}$). Plasma in these conditions couples with radiation
mainly through electronic transitions of elements
(C, N, O) in their He-like and H-like ionization stages.
The strongest of these transitions falling in the soft X-rays.  Due to the extreme low densities and small optical
length of the CGM, detection of the absorption/emission
features produced by this material has proved particularly
challenging with the still limited sensibility of the
gratings on board Chandra and XMM-Newton. As a
consequence, observations in only a few number of line
of sight directions have been useful to observe X-ray
absorption in quasar spectra. In this scenario it is clear
that simulations can play an important role in the study
of hot gas as they let us to explore the complexities in
a simulated galactic systems. Including such information
enable us to study how observational techniques can
lead to biased results for the total hot gas in the galactic
system.\\
With such aim we have located our mock
observer at 8~kpc from the galactic center (i.e. similar
to solar position) in the galactic plane and at a random
azimuthal angle, inside our MW-sized simulations. For
comparison we refer the reader to observations of OVII
column densities presented in \citet{Gupta2012} and to total hot gas values they obtained. In Tab.~\ref{tab:4} we show the
 results we have obtained when making observations to a similar number of randomly distributed directions into the sky as the ones in \citet{Gupta2012}.
As in our simulations we have not realistic chemical species like
oxygen we have assumed that we are observing all hot gas, and present the column densities measured in each line of sight.
As it is common in the literature \citep[e.g. ][]{Gupta2012,Bregman&Lloyd-Davies2007}, we have assumed a spherical uniform gas distribution for
 the hot gas halo. As a first attempt, we have estimated, from the column density values at each direction, the total galactic hot gas
mass, assuming a path length of L~=~239~kpc~=~r$_{vir}$. We have found 
that using the spherical uniform approach, the total hot gas mass is systematically overestimated (see Tab.~\ref{tab:4} last column). 
This is a consequence of that the density distribution in our model is not uniform but it decreases as a power-law (see Fig.~\ref{fig:5}
and \ref{fig:dens}).
 However, this calculation is not very useful, as it requires an a-priori knowledge of the CGM path length (L) which is 
observationally unknown.\\

A popular technique among observational studies \citep[i.e.][]{Gupta2012,Bregman&Lloyd-Davies2007} is to take advantage
of the two observables for the hot gas in the CGM: the column density (N$_H$) and the emission measure (EM). This technique uses
 these two quantities to derive the total hot gas mass with no need of imposing an a-priori mean density or optical path length.
 We show results from using this 
technique in Tab.~\ref{tab:3}, both for the optical path length, the mean density and the total hot gas mass. In this case, as can
be seen in the table, results show an underestimation of the total hot gas mass by factors $\sim$0.7$-$0.1, result that is independent
 on the observed direction.
Again this result is a consequence of that the real density distribution of our model is not uniform but decreasing with radius as a 
power law.\\
As a conclusion we state that using both, column density plus imposed optical depth (independently of using the value from a single
 observation or the mean) or a combination of column density and emission 
measure, we are not able to obtain the real hot halo gas mass, when assuming it is isotropically distributed in the 
galactic halo. Nevertheless, these measurements, particularly the observational method in the literature  using both 
N$_H$ and EM are useful to get an order of magnitude estimate of the total hot gas mass of the halo.
The work is in progress to find the density profile that will allow us to get the best hot halo gas mass estimation, from
observations.

\begin{table}
    \tabcolsep 4.pt
\begin{tabular}{lcccc}
\hline
\hline
           &  l     &  b     & N$_H$                       &  M$_{total,Hotgas}$   \\
           & [deg]  &[deg]   & [10$^{19}$cm$^{-2}$]     & [10$^{10}$ M$_{\odot}$]\\
\hline
FoV1        & 179.83 & 65.03  & 2.59                  &   4.57                \\
FoV2        & 17.73  & -52.25 & 1.31                   &  2.31                \\
FoV3        & 91.49  & 47.95  & 2.96                  &   5.21               \\ 
FoV4        & 35.97  & -29.86 & 2.81                &     4.96             \\
FoV5        & 289.95 & 64.36  & 2.62                  &   4.62               \\
FoV6        & 92.14  & -25.34 & 2.06                   &  3.63                \\
FoV7        & 287.46 & 22.95  & 2.62                   &  4.62                \\
FoV8        & 40.27  & -34.94 & 2.59                  &   4.57               \\
\hline
\hline
\end{tabular}
\centering
\caption{Mock observations of eight randomly distributed directions in the sky, in our G.321 simulation at z~=~0. The total hot gas mass in the simulation, up to r$_{vir}$, is M$_{vir}$~=~1.2$\times$10$^{10}$~M$_{\odot}$. 
We have defined hot gas as gas at T~$>$~3$\times$10$^5$~K. We have assumed L~=~230.1~kpc~=~r$_{vir}$.}
\label{tab:4}
\end{table}

\begin{table}
    \tabcolsep 4.pt
\begin{tabular}{lcccc}
\hline
\hline
            & EM                       & L     & n                    & M$_{total,Hotgas}$   \\
            & [10$^{-3}$cm$^{-6}$~pc]  & [kpc] & [10$^{-4}$cm$^{-3}$] & [10$^{10}$ M$_{\odot}$]\\
\hline
FoV1        & 2.03                     & 40.2  & 2.2                  & 0.11 \\
FoV2        & 0.16                   &  111.3 & 0.4                  & 0.40 \\
FoV3        & 1.58                  &   53.7  & 1.7                  & 0.21 \\ 
FoV4        & 0.79                &     56.6  & 1.2                  & 0.17 \\
FoV5        & 0.89                  &   80.5  & 1.1                  & 0.42 \\
FoV6        & 1.40                   &  59.1  & 1.5                  & 0.25 \\
FoV7        & 0.98                   &  74.1  & 1.1                  & 0.36 \\
FoV8        & 0.61                  &   116.5 & 0.7                  & 0.88 \\
\hline
\hline
\end{tabular}
\centering
\caption{Emission Measure (EM) values from the same eight randomly distributed directions in the sky as in Tab.~\ref{tab:4}.
Using EM and N$_H$ values we have computed the path length (L) and hot gas volume density (n). Finally we show the total hot gas mass
 obtained when assuming a sphere with constant density (n) and its radius equal to the path length (L). 
The real total hot gas mass in the simulation is M$_{vir}$~=~1.2$\times$10$^{10}$~M$_{\odot}$.}
\label{tab:3}
\end{table}

\subsubsection{Total hot gas mass: accounting for the missing baryons}

The baryonic fraction in our simulations corresponding
only for stellar and cold gas mass, falls in between
0.084 and 0.096. These numbers are far from the cosmic
baryon fraction  F$_{b,U}$=0.171$\pm$0.009 \citep{Dunkley2009,Planck2013}  but are comparable
to reported values for isolated spiral galaxies based on
weak lensing and stellar mass are
0.056  \citep{Hoekstra2005}. Such mismatch between cosmic and galaxy baryon
budget suggests that either gas has escaped halos or never reached them and it
and it is located in the Large Scale Structure warm-hot intergalactic 
(WHIM) gas \citep{Cen&Ostriker1999} or that warm-hot gas embedded in 
dark matter halos may account  at least for part of the
missing baryons if not all.
In our simulations the baryonic fraction reaches (0.107$-$0.120) 
even after adding the hot gas mass component inside halos. 
Because we do not include the feedback effect of an AGN, that may eject the gas outside
the halo, we can consider  our estimates 
as an upper limit to the halo hot gas mass.  
Based on the previous discussion we argue that the missing baryons 
must be placed  both like hot gas mass inside dark matter halos \citep[e.g.][]{Gupta2012,Miller&Bregman2013} and also in the IGM, along the filaments and far from the main galactic systems \citep{Rasheed2010,Eckert2015}. This is consistent  with recent studies for CGIM
in M31 \citep{Lehner2015} as well as gas stripping in LMC \citep{Salem2015}.\\

\section{Conclusions}\label{sec:conclusions}

In this paper we have introduced a new set of MW-sized cosmological N-body plus hydrodynamics simulations. Their resolution and realism are high enough 
to allow us to study stellar kinematics \citep{RocaFabrega2013,RocaFabrega2014} and evolution of large scale structures of these kind of galaxies. 
Most stellar disk parameters 
in our G.3 models are in agreement with observational ranges for the MW. Disk scale length and scale height as well as their dependence on the stellar 
age are reproduced. The simulated disk shows a large variety of large scale structures such
us spiral arms, a ring and a bar. It also shows a flared and a warped configuration.
The quality of our runs is comparable to the most recent works presented by \citet{Guedes2011} and \citet{Mollitor2014},
a detailed comparison can be seen in Table~\ref{tab:1}. Our simulations have a similar or a
larger number of resolution elements (stellar particles or gas cells), 
higher spatial resolution, and a smaller mass per particle than the ones in those works.
In addition, as we have shown in Fig.~\ref{fig:4}, our circular velocity curves are in a very good agreement with recent observations.
On the other hand, 
a spheroid (bulge+halo) that is 2$-$3 times more massive than the one observed in 
the real MW is present in our G.3 models. 
Since the density of this spheroidal component is relatively low in the central region of our galactic systems 
it does not drive to centrally peaked circular velocity curves.

We have also analyzed the SFH in our simulations. We have found that our SFR values at z=0 are lower than recent observational values. Also the
 SFH peak in our G.3 models occurs at higher redshift than the one predicted by
\citep{Behroozi2013}. In the most recent and realistic MW-sized simulations \citep{Guedes2011,Mollitor2014,Bird2013}, 
the predicted SFH is also not well reproduced. We have also found that total gas mass and its distribution in temperature
 slightly differ from MW observations. This mismatch
 in SFH and mass gas distribution is usual in simulations due to the lack of understanding of subgrid physics and to the difficulty of accounting for all physical processes involved in the formation and
evolution of galaxies. Studies like ours will help to constrain such subgrid models. Recently, it has been proposed 
that the inclusion of a kinematic feedback might solve the early star formation issue and thus change the position of the SFH peak.

The work presented here is
focused in the study of 
the distribution and the amount of potentially X-ray emitting gas (T~$>$~3$\times$10$^5$~K) of
our simulations. The comparison of the properties of the
hot gas in our simulation with the observations of the hot halo X-ray corona of
the MW and external galaxies was made by producing hot gas column density mock observations. 
Conclusions on this part follows:
\begin{itemize}
 \item Hot gas is not distributed homogeneously around the galactic halo. 
 \item There is no evidence of a hot gas thick disk in our models.
 \item By using the real mass of warm-hot
 gas in our simulations we conclude that about 53\%$-$80\% of the missing baryons can be accounted by hot gas in the halo corona. The exact value depends on the simulated galactic 
system (e.g. halo mass). We argue that the lacking baryons are placed in filaments (IGM), not inside main galactic systems, as preheated warm-hot gas \citep{Rasheed2010}. It is important 
to mention that in our simulations AGN feedback from a central galactic black hole has not been implemented. It is still open to debate how a relatively low mass black hole, as
the one in the real MW, could affect the mass and temperature distribution of the hot halo gas and thus the SFH.
 \item Hot gas density follows several streams in the halo that can be detected because of their changes in kinematics and metallicity. We have seen how low metallicity streams come from the
 IGM while others with higher metallicity depart from the disk. We have also detected the infall of high metallicity gas that can be interpreted as coming from a satellite accretion.
 \item The total hot gas mass in the halo can not be recovered when assuming a spherical uniform gas distribution. 
In our model the real hot gas density distribution lead to an overestimation when using the
mean column density plus a fixed optical depth under the spherical uniform assumption. On the other hand it drives to an underestimation of the total hot gas mass
when we use the column density plus the emission measure when using the same assumption.
 \item Following previous works we have found that a clear relation exist between total hot gas mass and virial mass of halos. This result is 
important as it becomes new method to constrain the total mass of galactic systems, such as the Milky Way.
\end{itemize}

\section*{Acknowledgments}

We thank A. Klypin and A. Kravtsov for providing us the numerical codes. We thank HPCC project, T. Quinn and N. Katz for the implementation of
TIPSY package.
This work was supported by the MINECO (Spanish Ministry of Economy) - FEDER through grant AYA2012-39551-C02-01 and ESP2013-48318-C2-1-R and Conacyt Fronteras de la Ciencia through grant 281.
The simulations were performed at the Supercomputer Miztli of
 the DGTIC-UNAM, and at Atocatl and Abassi2, HPC facility of the Institute of Astronomy-UNAM.


\bibliographystyle{mn2e}

\bibliography{biblio}
\IfFileExists{\jobname.bbl}{}
{\typeout{}
\typeout{****************************************************}
\typeout{****************************************************}
\typeout{** Please run "bibtex \jobname" to optain}
\typeout{** the bibliography and then re-run LaTeX}
\typeout{** twice to fix the references!}
\typeout{****************************************************}
\typeout{****************************************************}
\typeout{}
}

\begin{appendix} 
\section{Resolution study} 
\begin{figure}
\centering
\includegraphics[scale=0.5]{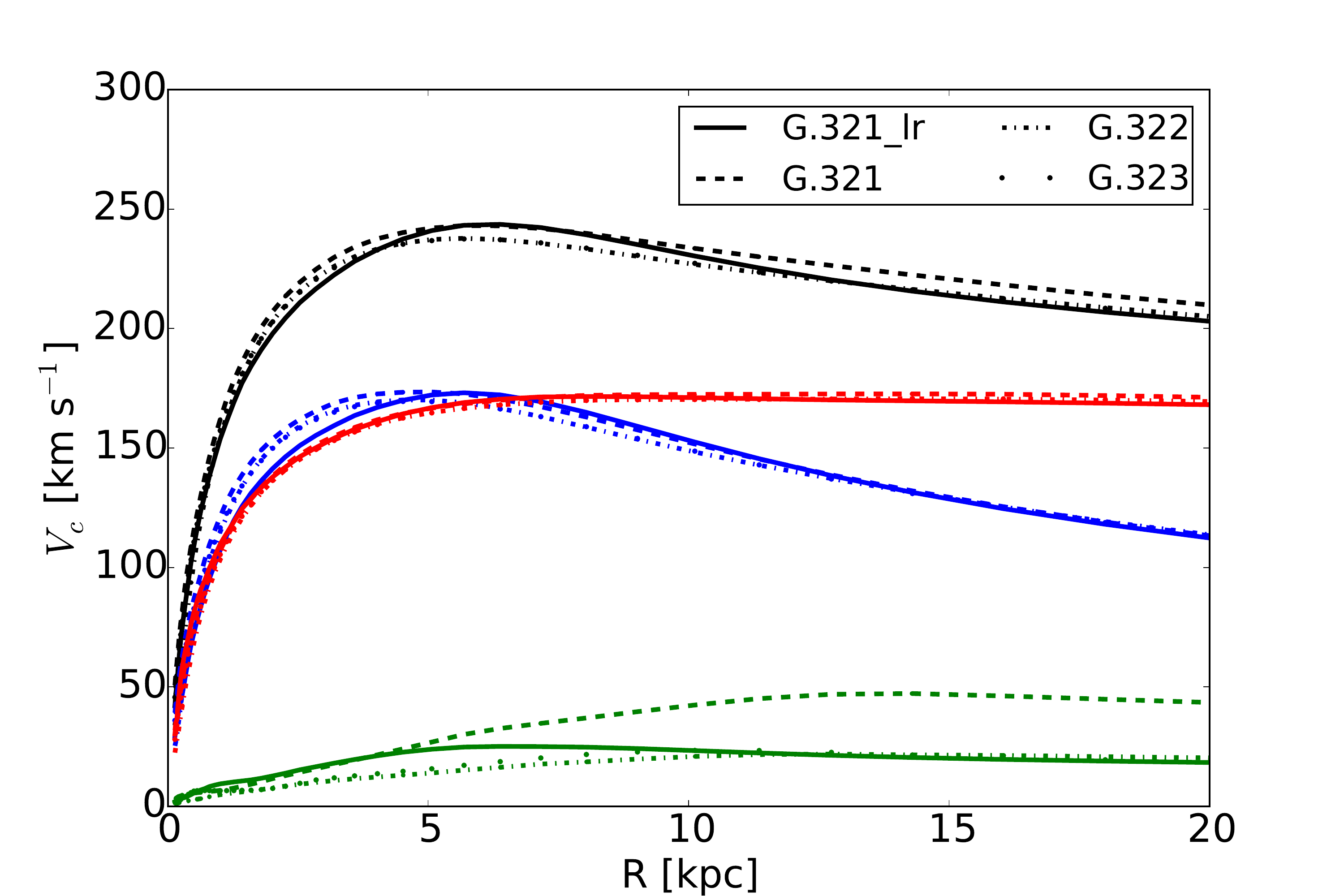}
\caption{Circular velocity curves of G.321, G.322, G.323 and G.321\_lr computed using GM/r approximation (see Tab.\ref{tab:App1} for more details about models). Black: total, red: dark matter, blue: stars, green: gas.}
\label{fig:A1}
\end{figure}

In this appendix we show a test of numerical convergence using a low resolution resimulation of our main model. The resimulation presented here (G.321\_lr) has been obtained using the same IC and initial parameters as in G.321 but with a resolution of 218 pc instead of 109 and $\sim$9$\times$10$^{5}$ DM particles inside R$_v$
instead of $\sim$7$\times$10$^{6}$. In Fig.\ref{fig:A1} we show a comparison of the total, DM, stellar and gas circular velocity curves computed using $\sqrt(GM/r )$ as a proxy for V$_c$. We see that resolution does not affect the total circular velocity curve profiles. However, it is interesting to note that circular velocity curve from the G.321 gas component shows significantly higher V$_c$ values at large radii. This result is in agreement with what was obtained by \citet{Scannapieco2012}; i.e. the properties of the gaseous component seem to  be the most sensitive to numerical resolution effects. We show in Tab.\ref{tab:App1} the values of the main parameters of the G.321 run and compare them with the corresponding ones of the G.321\_lr model. We see that, aside from the mass of cold gas and related quantities, like SFR at z=0,  most parameters do not differ from each other by ore than 30$\%$. This was also found in \citet{Scannapieco2012}. From this result, we conclude that although, in general, parameters are not sensitive to changes in resolution, caution must be exercised when analyzing the cold gas component and the SFR. In addition, we notice that the cold gas fraction is not only sensitive to numerical resolution (the maximum refinement level) but also to changes in terms of number of cells (different aggressive refinement, see Sec.3.3). On the other hand, the difference observed in G.321 gas velocity curve at large radii is simply due to the fact that this run ends up with much more gas. Additional studies of numerical convergence of the code can be found in \citet{AvilaReese2011,GonzalezSamaniego2014a}.

\begin{table}
    \tabcolsep 4.pt
\begin{tabular}{|lcc|c|}
\hline
\hline
                                & G.321           &  G.321\_lr  & $\mid$(G.321-G.321\_lr)/G.321$\mid$$\times$100 \\
\hline
M$_{vir}$ [M$_{\odot}$]         & 7.33$\times$10$^{11}$    & 7.33$\times$10$^{11}$ & 0.0 \\
r$_{vir}$ [kpc]                 & 230.1                            & 230.5       &  0.2          \\
M$_{*}$ [M$_{\odot}$]           & 6.1$\times$10$^{10}$     & 6.0$\times$10$^{10}$ & 2.0 \\
M$_{gas}$ [M$_{\odot}$]         & 2.70$\times$10$^{10}$    & 1.93$\times$10$^{10}$ & 28.5\\
M$_{hotgas}$ [M$_{\odot}$]      & 1.22$\times$10$^{10}$    & 1.06$\times$10$^{10}$ & 13.1 \\
M$_{warmgas}$ [M$_{\odot}$]     & 5.66$\times$10$^{9}$      & 7.33$\times$10$^{9}$ & 29.5\\
M$_{coldgas}$ [M$_{\odot}$]     & 9.34$\times$10$^{9}$     & 1.37$\times$10$^{9}$ & 85.3\\
M$_{200}$ [M$_{\odot}$]         & 6.84$\times$10$^{11}$    & 6.61$\times$10$^{11}$ &  3.4\\
r$_{200}$ [kpc]                 & 175.6                                   & 175.6 & 0.0\\
F$_{b,U}$                       &  0.120                                & 0.108  & 10.0\\
c                               &  28.5                                  & 25.4 & 10.9\\
R$_{d}$ [kpc]                   &  2.56 (4.89/2.21)          & 2.13 (4.2/1.8) & 16.8\\
h$_{z,young}$ [pc]              &  277                                     & 390 & 40.7\\
h$_{z,old}$ [pc]                &  1356                                  & 1032 & 23.9\\
$\alpha_X$                      &  -0.62                            & -0.60 & 3.2 \\
SFR (z=0) [M$_{\odot}$yr$^{-1}$]&  0.27                         & 0.16 & 40.7\\
V$_{c\odot}$(R=8kpc) [kms$^{-1}$] &  239.8                        & 239.2 & 0.3\\
R$_{peak}$ [kpc]                &  5.69                              & 5.69 & 0.0 \\
V$_{c}$(R$_{peak}$) [kms$^{-1}$]   &  243.8                            & 243.6 & 0.1\\
V$_{2.2}$/V$_{200}$             &  1.90                                  & 1.87  & 1.6\\
\hline
\hline
\end{tabular}
\centering
\caption{Comparison between general parameters of GARROTXA original model G321 and low resolution run G321\_lr. All parameters shown here have been well described in Tab.\ref{tab:1}.}
\label{tab:App1}
\end{table}
\newpage

\section{Jeans length} 

\begin{figure}
\centering
\includegraphics[scale=0.45]{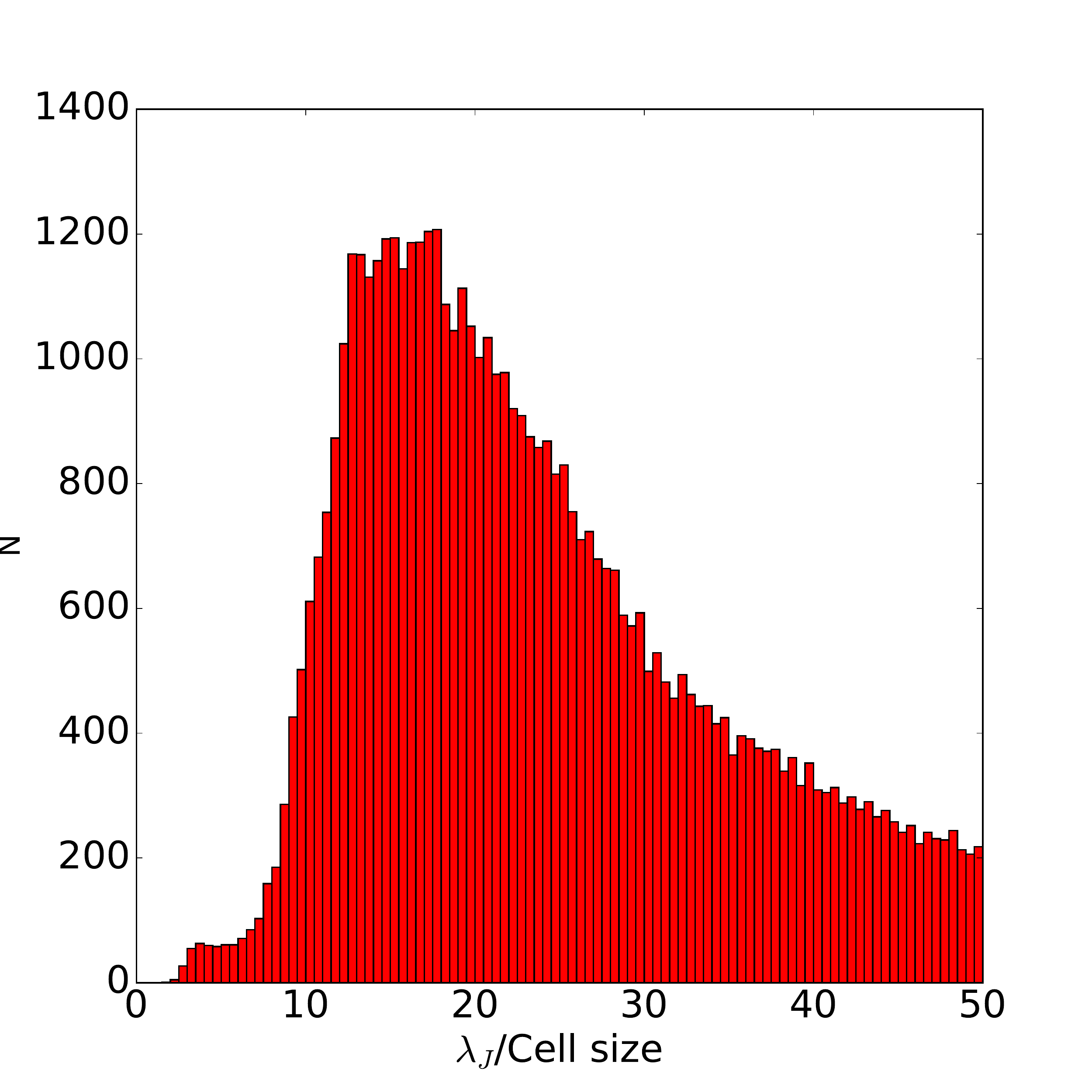}
\caption{Histogram of the cold gas (T$<$4$\times$10$^4$) Jeans length computed following Eq.\ref{eq:1}, in units of the corresponding cell size, for simulation G322.}
\label{fig:B1}
\end{figure}

In this section, following the condition proposed by \citet{Truelove1997}, we show that in our simulations the resolution is above the minimum that ensures the formation of stellar particles is of physical origin rather than numerical. In order to see this we have computed the local Jeans wavelength  of the cold gas ($\lambda_J$) according to Eq.\ref{eq:1}.\\
In Fig.\ref{fig:B1} we show a histogram of the Jeans length in units of the size of the corresponding cell for simulation G322. Except for few ones, cells satisfy the Truelove criterium.\\

\end{appendix}

\def\apj{ApJ}
\def\apjl{ApJ}
\def\aj{AJ}
\def\mnras{MNRAS}
\def\aa{A\&A}
\def\nat{nat}
\def\araa{ARA\&A}
\def\aap{A\&A}



\label{lastpage}
\end{document}